%% file: rsc-articletemplate-softmatter.tex
\documentclass[twoside,twocolumn,9pt]{article}
\usepackage[final]{changes}
\usepackage{extsizes}
\usepackage[super,sort&compress,comma]{natbib} 
\usepackage[version=3]{mhchem}
\usepackage[left=1.5cm, right=1.5cm, top=1.785cm, bottom=2.0cm]{geometry}
\usepackage{balance}
\usepackage{mathptmx}
\usepackage{sectsty}
\usepackage{graphicx} 
\usepackage{lastpage}
\usepackage[format=plain,justification=justified,singlelinecheck=false,font={stretch=1.125,small,sf},labelfont=bf,labelsep=space]{caption}
\usepackage{float}
\usepackage{fnpos}
\usepackage[english]{babel}
\addto{\captionsenglish}{%
  
}
\usepackage{amssymb}
\usepackage{array}
\usepackage{droidsans}
\usepackage{charter}
\usepackage[T1]{fontenc}
\usepackage{setspace}
\usepackage[compact]{titlesec}
\usepackage{hyperref}

\usepackage{epstopdf}
\usepackage{multirow}
\usepackage{float}
\floatstyle{plaintop}
\restylefloat{table}

\definecolor{cream}{RGB}{222,217,201}

\begin{document}

\definechangesauthor[color=magenta]{R2}

\thispagestyle{plain}




\makeFNbottom
\makeatletter
\renewcommand\LARGE{\@setfontsize\LARGE{15pt}{17}}
\renewcommand\Large{\@setfontsize\Large{12pt}{14}}
\renewcommand\large{\@setfontsize\large{10pt}{12}}
\renewcommand\footnotesize{\@setfontsize\footnotesize{7pt}{10}}
\makeatother

\renewcommand{\thefootnote}{\fnsymbol{footnote}}
\renewcommand\footnoterule{\vspace*{1pt}%
\color{cream}\hrule width 3.5in height 0.4pt \color{black}\vspace*{5pt}} 
\setcounter{secnumdepth}{5}

\makeatletter 
\renewcommand\@biblabel[1]{#1}            
\renewcommand\@makefntext[1]%
{\noindent\makebox[0pt][r]{\@thefnmark\,}#1}
\makeatother 
\renewcommand{\figurename}{\small{Fig.}~}
\sectionfont{\sffamily\Large}
\subsectionfont{\normalsize}
\subsubsectionfont{\bf}

\setlength{\skip\footins}{0.8cm}
\setlength{\footnotesep}{0.25cm}
\setlength{\jot}{10pt}
\titlespacing*{\section}{0pt}{4pt}{4pt}
\titlespacing*{\subsection}{0pt}{15pt}{1pt}


\makeatletter 
\newlength{\figrulesep} 
\setlength{\figrulesep}{0.5\textfloatsep} 

\newcommand{\topfigrule}{\vspace*{-1pt}%
\noindent{\color{cream}\rule[-\figrulesep]{\columnwidth}{1.5pt}} }

\newcommand{\botfigrule}{\vspace*{-2pt}%
\noindent{\color{cream}\rule[\figrulesep]{\columnwidth}{1.5pt}} }

\newcommand{\dblfigrule}{\vspace*{-1pt}%
\noindent{\color{cream}\rule[-\figrulesep]{\textwidth}{1.5pt}} }

\makeatother

\twocolumn[
  \begin{@twocolumnfalse}

\vspace{1em}
\sffamily
\begin{tabular}{m{4.5cm} p{13.5cm} }


\includegraphics{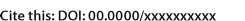} & \noindent\LARGE{\textbf{Subtleties of UV-crosslinking in microfluidic particle fabrication: UV dosage and intensity matter$^\dag$}} \\
\vspace{0.3cm} & \vspace{0.3cm} \\

 & \noindent\large{Sabrina Marnoto,\textit{$^{a}$} Avi J. Patel,\textit{$^{a}$} and Sara M. Hashmi$^{\ast}$\textit{$^{abc}$}} \\

\includegraphics{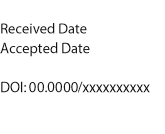} & \noindent\normalsize{Curable hydrogels have tunable properties that make them well-suited for applications in drug delivery, cell therapies, and 3D bioprinting. Advances in microfluidic droplet generation enable rapid fabrication of polymer-filled droplets. UV-curable polymers offer a clear path toward using fluidic generation to produce monodisperse microgels with uniform properties. In flow, polymer concentration and UV exposure both control the degree of crosslinking. High UV intensity is often used to ensure complete gelation and avoid complications that may arise from partial curing.  Optical microscopy can assess droplet and particle sizes in flow. However, optimizing formulations for mechanical properties usually requires removal of generated material and external measurement outside of flow. In this study, we couple droplet generation, microgel fabrication, and mechanics assessment within a single fluidic device.  We make and measure soft polyethylene glycol diacrylate (PEGDA) microgels by curing polymer-filled water drops in mineral oil. Crosslinking is tuned by varying UV dosage, allowing us to study how gelation degree influences microgel properties. Within the device, we use shape deformation in flow to measure the restoring stress of both droplets and particles. Our results suggest that PEGDA droplets gel from the inside out. If gelation is incomplete, a particle resides within a fluid drop.  Independent measurements outside of flow corroborate this observation. Crosslinking PEGDA-filled droplets in a pendant drop geometry, with dye, suggests the persistence of an aqueous shell around the gel. Similarly, microparticles in PEGDA-filled drops undergoing gelation exhibit diffusive arrest near the drop center, while maintaining mobility in an outer region.  Together, these results suggest the importance of considering the extent of gelation when fabricating microgels using fluidics.}

\end{tabular}

 \end{@twocolumnfalse}
 
 ]

\vspace{0.6cm}


\renewcommand*\rmdefault{bch}\normalfont\upshape
\rmfamily
\section*{}
\vspace{-1cm}

\footnotetext{\textit{$^{a}$~Department of Chemical Engineering, Northeastern University, Boston, MA 02115, USA.}}

\footnotetext{\textit{$^{b}$~Department of Mechanical Engineering, Northeastern University, Boston, MA 02115, USA.}}

\footnotetext{\textit{$^{c}$~Department of Chemistry and Chemical Biology, Northeastern University, Boston, MA 02115, USA. Email: s.hashmi@northeastern.edu}}
\footnotetext{$\ast$~Email: s.hashmi@northeastern.edu.}



\section{Introduction}
From repairing microcracks in essential coated fabrics to sustaining long-term drug release, soft UV-curable hydrogel microparticles offer innovative solutions for various applications.\cite{gong2024,xue2015}  Among the most popular applications for hydrogels results from their biocompatibility and  tunable mechanical properties.  Hydrogels are used as scaffolds for a variety of biological materials from single cells to tissues \cite{el2013hydrogel, farasati2024hydrogel}.  The tunable mechanics of hydrogels allows them to mimic the extracellular matrix, ensure cell viability, and even control stem cell differentiation. \cite{tibbitt2009hydrogels, guvendiren2010control, benoit2008small} To encapsulate individual cells, fluidic methods are often used to ensure a high degree of uniformity and enable a variety of downstream assays in flow. \cite{dendukuri2009synthesis}
 
Microfluidics enables the high-throughput production of monodisperse particles by first using a droplet maker to generate droplets containing polymer.  Crosslinking into a gel can be accomplished by the addition of a crosslinker, as is common with gels like alginate \cite{aguilar2021formation, liu2006shape}, or inclusion of a photo-initiator. Droplets are then cured either in flow or as they exit the channel and are removed entirely from the chip. Off-chip curing techniques, especially those involving photo-crosslinking, are popular. \cite{kim2013,PULLAGURA2021, ChenandAluunmani2022}  However, off-chip curing negates a main advantage of microfluidics: the opportunity for continuous production followed by in-line, downstream analysis. Further, droplets can coalesce as they exit fluidic devices into larger reservoirs, nullifying the monodispersity advantage of fluidic techniques. With transparent or UV-transmittable device materials, in-device photo-polymerization is possible by exposing a region of the fluidic device to UV light.\cite{Seiffert2011,wu2022} On-chip polymerization is already being implemented in several microfluidic platforms. \cite{Hinojosa-ventura2023, Akbari2017, wu2022}  
 
In the context of photocrosslinking, the use of on- or off-chip approaches assumes that particles are fully cured after either application of high UV intensity or long exposure time.\cite{PULLAGURA2021, Anselmo2015, xue2015, rekowska2023} Although complete gelation is often assumed, both exposure time and UV intensity uniquely tune properties such as size, shape, and structure. \cite{Filatov2017, ChenandAluunmani2022,Jiang2021} Precision in UV curing also avoids instances of over- or under-curing. In some cases, partial microparticle curing may be desired. For example, UV-curable micro and nanoparticle surfaces can improve cell adhesion using partial UV curing.\cite{Li2013} Partial curing has also been exploited to permanently reconfigure microparticles into more desirable shapes for photonic and drug carrier applications.\cite{cox2017} These examples highlight the importance of precise UV curing control to tailor microparticles to specific functions. 




Several experimental methods can quantify gelation dynamics, both on micro and macro scales. Bulk rheology coupled with UV illumination provides unique insights into the dynamics of gelation and evolution of material properties.\cite{burroughs2022gelation, bonino2011real}. Long standing methods like photo-differential scanning calorimetry are still used, for instance to extract thermal effects of UV-cured tripropylene glycol diacrylate monomers.\cite{hadjou2019} On a microscale, a time-resolved Fourier transform infrared technique measures the double bond consumption rate as microgels form in a UV curing reaction.\cite{nebioglu2006} Microscopy and visual techniques also monitor UV crosslinking.\cite{Filatov2017,Jiang2021} While a multitude of metrics exist to infer gelation dynamics, few studies compare them simultaneously with precise UV control and flow conditions for materials made by fluidic means.

In this paper, we use poly(ethylene glycol) diacrylate (PEGDA) as a model hydrogel system to investigate crosslinking both inside and outside a microfluidic platform. We explore a wide array of methods to measure PEGDA gelation and the resulting change in mechanical properties, including measurements of shape deformation in flow, nanoparticle diffusion measurements, and visual confirmation of gelation both within and outside a fluidic device. From the combination of these three methods, we obtain a comprehensive understanding of dynamic particle gelation both in and out of flow. Our results provide guidelines for researchers in designing microparticle curing protocols.

\section{Materials and Methods}
\subsection{Solutions}
We prepare solutions of poly(ethylene glycol) diacrylate (PEGDA) of molecular weight 700 \added{and density of 1.12 g/mL}(Sigma-Aldrich, CAS 26570-48-9), adding a photoinitiator, 2-hydroxy-2-methylpropiophenone (Sigma-Aldrich, CAS 7473-98-5, \added{98\% purity}).  Solutions are prepared at $c=9$, $19$, $29$, and $39$\% by volume PEGDA in DI water, all with a constant $1\%$ by volume of photoinitiator.  \added{The maximum density of the PEGDA solutions is 1.05 g/mL.} Table \ref{tab: PEGDA Viscosity Table} below presents the viscosity, $\mu$, of pertinent PEGDA solutions before UV exposure, measured using a Rheometer (TA DHR; $2^0$, 60 mm cone) \added{using a shear sweep from 1 to 100 1/s.  The aqueous PEGDA solutions are all Newtonian}.


Emulsions are prepared by mixing aqueous PEGDA solutions with mineral oil (Fisher, CAS 8042-47-5) and sorbitan monooleate (Span 80) (Sigma-Aldrich, CAS 1338-43-8) at mole fraction $\chi=10^{-2}$. \added{Mineral oil has a density of 0.83 g/mL and viscosity of 40 mPa-s. The viscosity is measured using the same protocol as above with a rheometer.} Three methods incorporate emulsions: bulk emulsification, a pendant drop setup, and fluidics. \added{All measurements are done at room temperature.}



\begin{table}[h!]
\centering
\begin{tabular}{|c|c|}
\hline
$c$ (\%) & $\mu$ (mPa-s) \\
\hline
0 & 0.89 \\ \hline
9 & 1.47 $\pm$ 0.00 \\ \hline
19 & 2.46  $\pm$ 0.20\\ \hline
29 & 4.23  $\pm$ 0.12\\ \hline
39 & 7.77  $\pm$ 0.31\\ \hline
\end{tabular}
\caption{Viscosity, $\mu$, of each relevant volume concentration, $c$, of uncrosslinked PEGDA solutions. Error bars represent three different viscosity experiments.}
\label{tab: PEGDA Viscosity Table}
\end{table}

\subsection{Fluidic Device Fabrication and Flow Set up}

Polydimethylsiloxane (PDMS) and silicone curing agent (Sylgard) mold microfluidic devices using a standard soft lithography technique \cite{mcdonald2000fabrication}. The PDMS is bound to a glass slide after both are treated with plasma (Harrick) and holes are punched into the inlets and outlets of the PDMS. We apply sunscreen (Cerave) on the upstream PDMS to ensure minimal crosslinking of the inlets. Aquapel (Debaishi) pretreatment of the devices makes the surfaces hydrophobic to prevent  sticking of the aqueous PEGDA solution.  Aquapel is injected into and fills the entire device, which is then heated overnight at 50\textdegree C. We remove the Aquapel by flushing the device with mineral oil containing $\chi=10^{-2}$ Span 80 before use.  

We use the two fluidic designs shown in Fig. \ref{UVschematic3}a and b. Both designs feature three inlets feeding a droplet maker, highlighted in yellow. Aqueous PEGDA solutions flow from inlet 1. Mineral oil with $10^{-2}$ Span 80 pinches off PEGDA droplets from inlet 2 and 3. Inlet 3 acts as a velocity control and ensures droplets do not collide or coalesce. The droplets then enter a long residence time channel, highlighted in purple, where they are exposed to UV light.\par

The first device design (\ref{UVschematic3}a) is either 50 or 70 $\mu$m tall with a 20 mm long channel located upstream of a series of stacked constrictions, shaded in blue.  The constrictions have widths that start from 200 $\mu$m and decrease in 40$\mu$m increments to the narrowest width of 40 $\mu$m. The 20 mm long channel yields residence times $t_r$ between 3.5 and 20.7 s, controlled by varying the droplet velocity. \added{Given that the maximum density of the droplets is 1.05 g/mL, the Stokes settling velocity is approximately 5 $\mu$m/s, which is several orders of magnitude slower than the droplet velocity. Further, residence times within the constriction region highlighted in blue in Fig. \ref{UVschematic3}a are $\sim O(1)$s.  As droplets traverse the microscopic imaging field of view, they may settle a few microns.  However, drop diameters are $\sim O(10) \mu$m, and any settling during measurement is negligible compared to their size. }\par

The second device, Fig. \ref{UVschematic3}b, is 100 $\mu$m tall and has a longer residence time channel, 56 mm long, which yields residence times of 6 to 40 s. The droplets or cured particles then pass through a series of constrictions and expansions, highlighted in blue. Seven constrictions in series decrease from 100 to 40 $\mu$m in width, in increments of 10$\mu$m.  After each constriction, the channel width returns to 500 $\mu$m.  \par

Fluid is driven at constant driving pressure (Fluigent LineUp Flow EZ pressure control system) for all fluidics measurements. Inlets 1, 2, and 3 are driven by pressures between 300 and 7000 mbar, with inlet 3 having the highest pressure.  \deleted{We calculate wall shear rate $\dot{\gamma}$ using $2v/H$, where $H$ is both the height of the channel and its narrowest dimension.} \par 

\subsection{Code/deformation characterization}
Either a sCMOS camera (Leica DFC9000) or a high-speed camera (Photron FASTCAM Mini AX200) captures videos of droplets moving through constrictions of both devices.  In each frame of the microscopy video, we use contour detection to identify both channel walls and the droplets or particles.  Standard particle tracking algorithms are used to identify trajectories based on centroid locations and then measure velocity.\cite{Crocker1996}  Contour detection enables measurement of both the unperturbed radius of a droplet or particle, $a$, and the elongation of deformed droplets or particles. In this way, our image analysis protocols determine droplet or particle size, instantaneous velocity, and deformation in each frame.


\subsection{UV calibration}

\begin{figure}[h!]
\centering
\includegraphics[width=1\columnwidth]{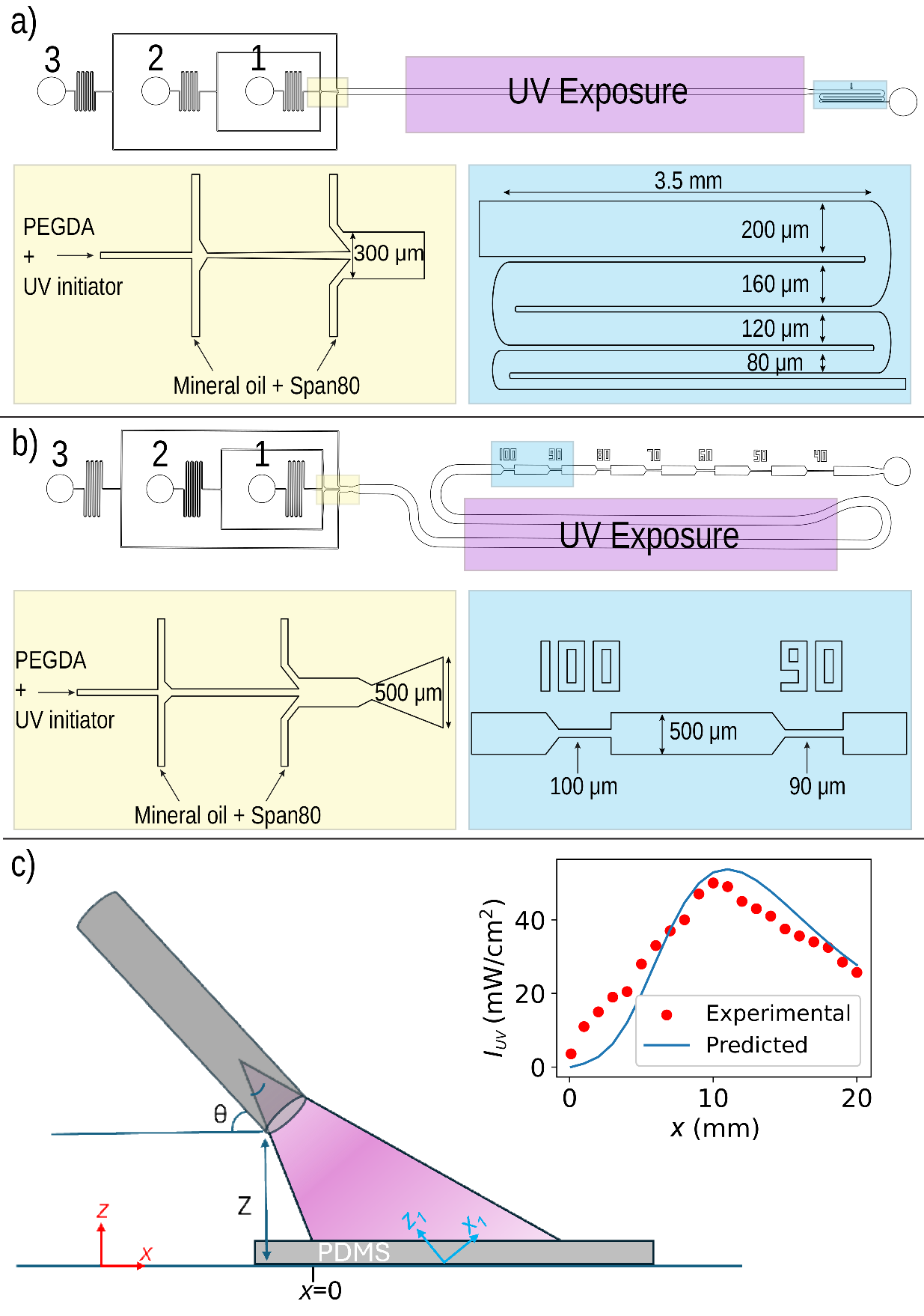}
\caption{Schematics of experimental setups. a and b show the two fluidic device designs. Both devices feature an upstream droplet maker and a long residence channel for UV exposure. The device in Fig. \ref{UVschematic3}a has a channel width which sequentially decreases from $200 \mu$m to $40 \mu$m, in intervals of $40 \mu$m.  The device in Fig. \ref{UVschematic3}b has a channel with width $500 \mu$m that is interrupted by a series of constrictions ranging from 100 $\mu$m to 40 $\mu$m. Panel c shows a schematic of the UV light setup: UV light shines from the tip of the spot curing light guide.  The UV light shines through the PDMS slab of the device to the droplet flow path on the glass slide. The inset to Fig. \ref{UVschematic3}c shows a calibration of equation \ref{eq: UV intensity final} to experimental values of $I_{UV}$ taken using a UV intensity meter.}
\label{UVschematic3}
\end{figure}

UV light from a 365 nm UV light guide (Lesco UV COOL CURE Lite) initiates free radical photo-polymerization between PEGDA and the photoinitiator, crosslinking the aqueous PEGDA solutions into hydrogels. Fig. \ref{UVschematic3}c shows a schematic of the light guide shining through a slab of PDMS onto the flow path. The axes are labeled in red. Droplets flow into the illuminated region along the $x$ axis and at constant $z=Z$, the height from the glass slide to the light guide, a constant for each flow test.  $x=0$ indicates the location of the initial exposure of the droplets to UV light.  The light guide is oriented at an angle $\theta$ from horizontal.  In the rotated coordinate system $(x_1, z_1)$, labeled in blue, $z_1$ lies along the axis of the beam.  Droplet position in this coordinate system is given by:
\begin{equation}
\begin{bmatrix}
x_1\\
z_1
\end{bmatrix} = \begin{bmatrix}
\sin{\theta} & -\cos{\theta}\\
\cos{\theta} & \sin{\theta}
\end{bmatrix} \begin{bmatrix}
x\\
Z
\end{bmatrix}
\label{eq: rotation matrix}
\end{equation}
Droplets experience a variation in UV intensity as they travel along the flow path. The UV light is a Gaussian beam whose width increases linearly with $R=(x^2 + Z^2)^{1/2}$, the distance from the light source. Intensity $I_{UV}$ decreases as the light spreads away from the light guide, following an inverse square law. In the rotated coordinate system, this intensity is given by:

\begin{equation}
I_{UV}(x_1, z_1)=  \frac{a R}{z_1^2}{\exp\left({\frac{-x_1^2}{(b + c z_1^2)^2}}\right)} NI_{UV_0}
\label{eq: UV intensity final}
\end{equation}


\noindent where $z_1$ is calculated using equation \ref{eq: rotation matrix} and $N$ is an integer indicating the control level on the light. We verify the linear relationship between $N$ and $I_{UV}$ using a UV intensity meter (Agiltron). We measure $I_{UV_0}$ at the exit of the light guide before every experiment.  PDMS does not strongly absorb UV light \cite{Seiffert2011}.  Measurements of UV intensity with and without PDMS between the light and the intensity meter confirm that the PDMS device thickness minimally affects UV intensity experienced in the flow path.  The inset to Fig. \ref{UVschematic3} shows the calibration of $I_{UV}$ as a function of $x$, using $N=1$, $\theta=36^{\circ}$, $z_1=14$ mm, and $I_{UV_0}=100$ mW/cm$^2$.  Fitting Eq \ref{eq: UV intensity final} to the measured light intensity provides $a=274$, $b=2.5$, and $c=0.16$.

For droplets in flow, UV energy dosage $\Omega$ depends on UV intensity, $I_{UV}$, and exposure time, $t_r$, the residence time of the droplet in the region of the channel shaded in purple in Fig.s \ref{UVschematic3}a and b. Both the intensity applied to the droplet and the exposure time of the droplet within the light path depend on its $x$-position. To calculate the total dosage for droplets in flow, intensity and exposure time are integrated over the droplet's trajectory in $x$:

\begin{equation}
\Omega= \int I_{UV}(x)t_r(x) dx   
\label{eq: dosage integral}
\end{equation}

We also calculate UV dosage in stationary contexts. For a stationary droplet or particle, $\Omega= I_{UV}t$. Fig. \ref{fig: SIschematic} shows experimental schematics of UV illumination for gelation kinetics, capillary micromechanics, and pendant drop measurements, each described next.



\subsection{Crosslinking kinetics}

To measure PEGDA gelation kinetics, we load PEGDA solutions with polystyrene fluorescent nanoparticles (Fluospheres, Invitrogen).  Nanoparticles are 500nm in diameter, loaded at 0.004$\%$ by volume, and fluoresce at 505nm (515nm emission).  These measurements are done in two types of samples: in a bulk aqueous macrogels and in large emulsion drops in mineral oil. 


To measure gelation kinetics in bulk macrogels, we use capillary action to draw aqueous PEGDA solutions into a rectangular glass capillary with cross section 0.05 mm $\times$ 1 mm and 50 mm long. Epoxy seals the ends of the capillary tube to minimize drift and ensure that tracer particle motion represents diffusive Brownian motion.  A 4 mm-thick PDMS slab covers the capillary so that the geometry of UV exposure mimics conditions within the microfluidic device, as in Fig. \ref{UVschematic3}c.  Optical microscopy videos are obtained at 63X magnification and a frame rate of 82 fps to enable particle tracking measurements of diffusive motion before UV exposure and arrest after UV exposure.

In a separate measurement, we seed large emulsion drops with these same tracer particles.  Emulsions are made by gently mixing 0.01\% aqueous solutions in mineral oil to form large droplets ($\sim10\mu$L) with $c=39$\% PEGDA. We place the emulsion on a microscope slide and allow the droplets to settle onto the glass slide.  We expose the sample to low intensity UV light, $I_{UV}<30$ mW/cm$^2$, and collect 30s of video microscopy of the tracer particle motion after 60s of UV exposure.  Particle tracking over the course of the video measures the diffusion constant of tracers in various regions of the cured droplet.  



\subsection{Shear Microparticle Modulus on Batch Particles}

We use capillary micromechanics to measure the shear modulus of microparticles prepared in a batch suspension \cite{wyss2010}. A pipette puller (Sutter) pulls 1 mm outer diameter glass capillaries down to $\thicksim$15 $\mu$m outer diameter.  After pulling, the inner diameter at the exit of the capillaries ranges from 8 to 20 $\mu$m. Particles are pushed through a tapered capillary at constant pressure. The shape of the particles at rest, as a function of the applied pressure, is used to measure the shear modulus.  

To prepare particles, an emulsion of 1\% PEGDA solution in mineral oil is gently mixed by hand to encourage formation of large, polydisperse droplets.  We expose 1 mL of the emulsion to various doses of UV light, by shining the light onto the sample resting in the bottom of a 10 mL centrifuge tube. In this way we can use equation \ref{eq: UV intensity final} to calculate $I_{UV}$, with $\theta=90^{\circ}$ and $Z=40$ cm, the distance from the tip of the light guide to the sample.  We expose emulsions with $c=9, 19, 29, 39$\% PEGDA in the aqueous phase to $I_{UV}=1, 6,$ and $159$ J/cm$^2$.  Table \ref{tab: cap micromech ItD} provides exact UV intensities and exposure times for each emulsion. After UV exposure, we dilute the sample another 5 times in mineral oil and use a 200 $\mu$m filter (PluriSelect) to remove larger particles and prevent clogging upstream of the tapered end of the capillary. 

We use constant pressure (Fluigent LineUp Flow EZ) to push the sample into the tapered capillary.  The sample flows into the tapered capillary with an initial driving pressure $p_0$.  Once a single particle is captured inside the tapered capillary tip at $p_0$, we collect an image of its shape, and then increase the applied pressure in increments of 100 mbar. In this way, each captured particle shape is measured at 10 different values of $p$. For all measurements, $p_0=100$ mbar and measurements are taken until reaching $p=1000$ mbar.

\subsection{Pendant Droplet Crosslinking}

We use a hanging pendant droplet setup to visualize droplet crosslinking in real time in emulsion drops at rest, with diameters $\sim O(1)$ mm.  We add a purple dye at a concentration of 1.5 mg/mL to the PEGDA solutions to enhance contrast between the crosslinked and uncrosslinked regions of the drop.  We obtain the purple color by mixing blue and red food dyes in equal parts (Food Club).  We choose purple after screening several color options to enhance the contrast between cured and uncured PEGDA.  

Dyed aqueous solutions with varying concentrations of PEGDA are dispensed from a blunt needle of 0.5 mm outer diameter into mineral oil with a Span 80 mole fraction of 10$^{-2}$ in a glass cuvette.  Droplets hanging from the needle are exposed to UV light from one side of the glass cuvette, as shown in Fig. \ref{fig: SIschematic}c.  We record videos the droplets as they are exposed to UV light using a phone camera (Google Pixel 8a) combined with a smartphone macro photography lens (APEXEL). Colors within the drop evolve during the course of UV exposure, but only when the UV initiator is present in the solution; this is discussed further below.  We collect videos of each hanging droplet starting at the time of UV exposure until there is no further change in its appearance.  \added{We also perform these experiments with the aqueous phase and/or the continuous mineral oil phase purged with nitrogen for two minutes before loading the cell.}


\section{Results and Discussion}
\subsection{Using deformation to measure the progress of PEGDA gelation in flow}



We use microfluidic droplet generators to form PEGDA-filled water droplets and expose them to UV light as they flow.  UV-crosslinking transforms the mechanical properties of the droplets: liquid droplets have different mechanical properties than their photo-cured counterparts.  To investigate the degree of gelation accomplished by UV exposure, we assess the mechanical response of the suspended phase in flow, downstream of UV exposure.  To do this, we measure the deformation of the suspended droplets or particles as they are subject to increasing shear rates, and thus, shear stresses.  The deformation of a droplet is governed by surface tension, while the deformation of a uniformly gelled particle is governed by its elastic modulus.

\begin{figure*} [h!]
\includegraphics[width=\textwidth]{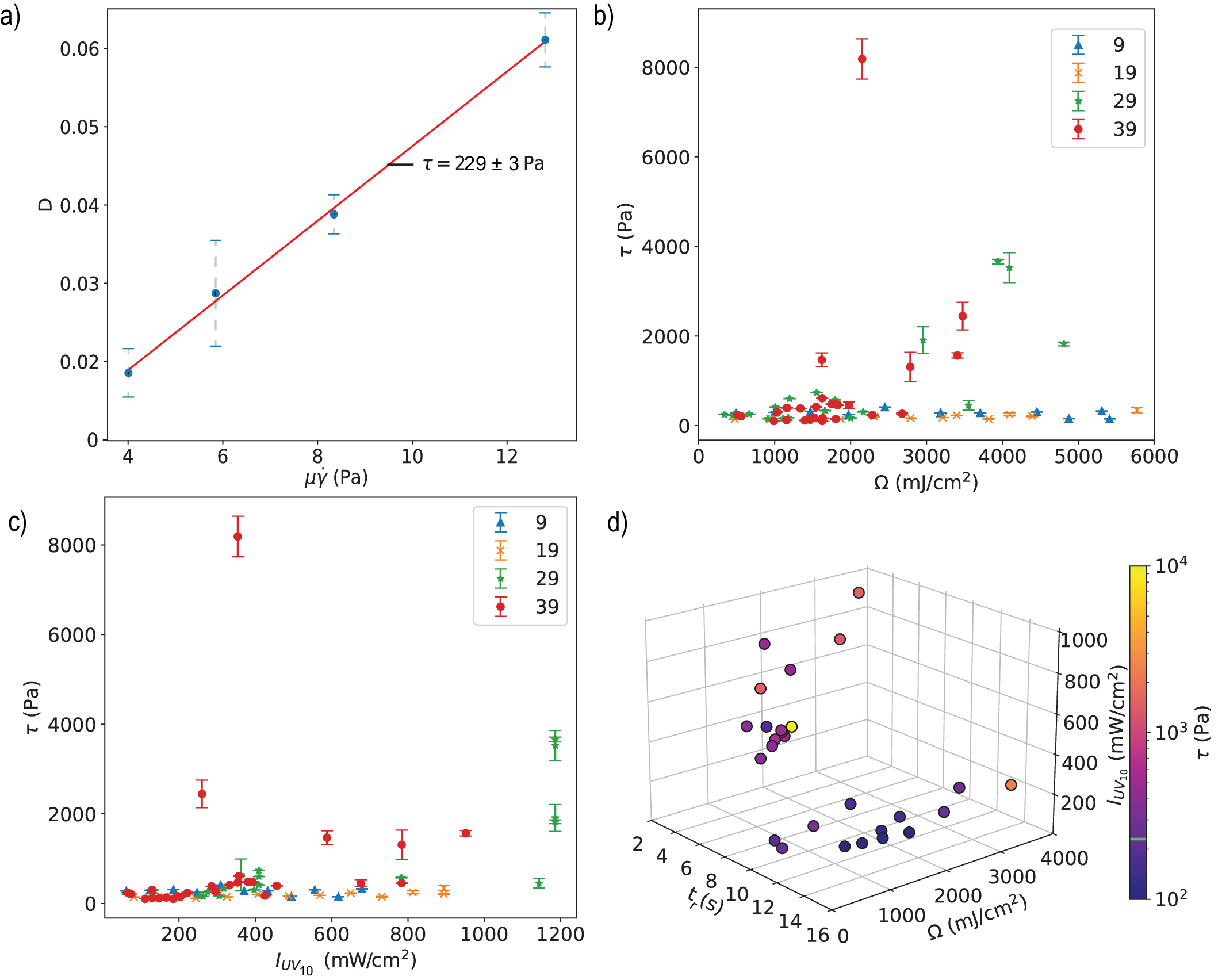}
\caption{Fig. \ref{fig: restoring stress}a: An example of the measured deformation in each constriction plotted as a function of viscous shear stress. Restoring stress, $\tau$ is the inverse slope of the linear fit. The \added[id=R2]{gray dashed} error bars on each data point represent the standard deviation for all \added[id=R2]{$\sim1000$} droplets or particles measured in each constriction. \added[id=R2]{The standard error of the mean is also plotted, in black, and is smaller than the size of each data point.} The error bars on the measurement of $\tau$ are given by the \deleted[id=R2]{\added{standard error of the}} \deleted[id=R2]{linear fit} \deleted[id=R2]{\added{corresponding to a 68\%}} \added[id=R2]{95\%} \added{confidence interval} \added[id=R2]{of the linear regression of the entire data set}. Fig. \ref{fig: restoring stress}b: Restoring stress as a function of UV dosage, $\Omega$, for PEGDA concentrations $9$ to $39\%$ as shown in the legend. Error bars represent error propagation from the linear fit in Fig. \ref{fig: restoring stress}a. Fig. \ref{fig: restoring stress}c shows the same data as in Fig. \ref{fig: restoring stress}b, plotted as a function of UV intensity at $x=10$ mm from the beginning of light exposure ($I_{UV_{10}}$). Fig. \ref{fig: restoring stress}d: A four-dimensional representation of $\tau$, indicated by the colorbar, as a function of $I_{UV_{10}}$, $\Omega$ and $t_r$ for $39$\% PEGDA. The green line on the colorbar represents $\tau=230$ Pa. } 
\label{fig: restoring stress}
\end{figure*}



In 1934, G. I. Taylor derived a relationship between the deformation of \deleted{a drop} \added{surfactant-free drops} in shear flow and the Capillary number, $\text{Ca}=\frac{\mu\dot{\gamma}a}{\sigma}$, where $\mu$ is the viscosity of the suspending fluid, \added{$\dot{\gamma}$ is the wall shear rate,} $a$ the radius of the unperturbed droplet, and $\sigma$ the surface tension. When subject to shear stresses in unbounded flow, droplets deform, as measured by a shape deformation parameter, $D$:
\begin{equation}
    D= \frac{L-B}{L+B}
\end{equation}
\noindent where $L$ and $B$ are the major and minor axes of the droplet.  Taylor determined $D= f(\lambda) \text{Ca}$.\cite{Taylor1934} The prefactor $f=\frac{19\lambda+16}{16\lambda+16}$ depends on the viscosity ratio $\lambda$, the ratio of droplet viscosity to continuous phase viscosity, and is $f\sim O(1)$.  Taylor's results hold for small droplet deformation, up to $\text{Ca}\sim0.2$ in steady shear and when $\text{Ca}<0.3$ in steady extensional flows.\cite{Taylor1934}  Taylor's theory for droplets in the absence of surfactant has been extended to descriptions of droplets covered partially and fully by surfactant.\cite{stone1990effects, sibillo2006drop, vlahovska2009small}  In these cases, the result remains that steady deformation increases linearly with $\text{Ca}$, with a slope $O(1)$, in both unbounded and bounded shear flows.  We have recently validated that the scaling relationship $D=\text{Ca}$ provides a reasonable measurement of droplet surface tension in confined, pressure-driven fluidic flows over a range of surfactant coverage and viscosity ratios \cite{Shaulsky2025}.  \added{We use this dimensional scaling argument in the small deformation limit by estimating $\dot{\gamma}$ using $2v/H$, where $H$ is both the height of the channel shown in Fig. \ref{UVschematic3}a and its narrowest dimension. This scaling approach requires corrections as deformation increases.  For instance, as $\text{Ca}$ increases above $\sim0.2$, droplets elongate into steady bullet shapes \cite{chattopadhyay2025motion}.}


For elastic particles deforming in steady shear flows, simulations reveal a similarly linear dependence of particle deformation on an effective Capillary number: $D= \frac{5}{4}\text{Ca}_e$. Here, $\text{Ca}_e = \frac{\dot{\gamma}\mu}{G}$ where $G$ is the bulk shear modulus of the particle.\cite{GaoandHu2009,Villone2014a,Villone2014b} Again, similar to the analysis for droplets, the pre-factor describing the relationship between $D$ and $\text{Ca}$ or $\text{Ca}_e$ is $O(1)$. 



In our experiments, we investigate the gelation transition from droplets to particles.  We measure $D$ as a function of shear rate $\dot{\gamma}$ as the material flows through a series of increasingly narrow constrictions, shown in Fig. \ref{UVschematic3}a in blue. To capture the transition from droplets to particles through the entire range of flow test conditions, we employ $D\thicksim\frac{\mu\dot{\gamma}}{\tau}$.  In this way the restoring stress $\tau$ is agnostic of the composition of the suspended material, regardless of whether it is a droplet, a fully-cured particle, or something in between those two limits.  For droplets, $\tau=\sigma/a$.

Fig. \ref{fig: restoring stress}a demonstrates this approach to measure $\tau$, showing $D$ plotted as a function of $\mu\dot{\gamma}$.  In the example shown in Fig. \ref{fig: restoring stress}a, $c=$ 39\% PEGDA and the illumination dosage is 558 mJ/cm$^2$.  Fig. \ref{fig: restoring stress}a shows four data points, representing flow through four constrictions.  However, Fig. \ref{UVschematic3}b shows five constrictions ranging from 200 to 40 $\mu$m.  In the example shown in Fig. \ref{fig: restoring stress}a, the average unperturbed particle radius is $a\sim25$ $\mu$m.  Because $2a$ is greater than the width of the narrowest constriction, we disregard deformation in the 40 $\mu$m constriction.  $D$ increases linearly with $\mu\dot{\gamma}$. 
The \added[id=R2]{gray dashed} error bars on \added{each} \deleted[id=R2]{\added{data point}} \added[id=R2]{value} of $D$ \added{in Fig. \ref{fig: restoring stress}a} represent the standard deviation \deleted[id=R2]{of the deformation} of all \added[id=R2]{$\sim1000$} particles \added{flowing} through \added{that particular} \deleted{each} constriction, \added[id=R2]{ranging between 6 and 24\% of the average.} \deleted[id=R2]{\added{We fit the data to a line and}} \added[id=R2]{Given the size of each data set, the standard error of the mean is $\sim30$ times smaller than the standard deviation: there is great confidence in the mean.  This standard error is also shown in Fig. 2a, as black error bars, all of which are smaller than the size of the data points.  We perform a linear regression of the entire data set representing all particles flowing through all four constrictions.} The inverse of the slope provides \deleted[id=R2]{$\tau = 210 \pm0.02$ Pa.} \added[id=R2]{$\tau = 229 \pm3$ Pa, corresponding to a 95\% confidence interval $\tau=[226, 232]$ Pa.} \deleted[id=R2]{\added{The error bars on $\tau$ come from the standard error of the linear fit.}} 
\deleted[id=R2]{\added{With measurements of $D$ at each value of $\mu\dot{\gamma}$ distributed normally}} 
\deleted[id=R2]{\added{around an average, the standard error of the linear fit $D(\mu\dot{\gamma})$ corre-}} 
\deleted[id=R2]{ \added{sponds to a 68\% confidence interval; a 90\% confidence interval is}} 
\deleted[id=R2]{\added{given by 1.65 times the standard error.}} 

Over the range of conditions investigated, deformation measurements range from as low as $D<0.005$ to $D\sim0.2$, over a range of shear rates $\dot{\gamma}\sim $ 57.9 to 703.4 1/s, with constant continuous phase viscosity $\mu=40$ mPa.s. \added{Measurements are obtained in the small deformation limit, with most values $D<0.1$. }

In our fluidic experiments, we cure PEGDA droplets at varying UV doses and intensities.  Fig.s \ref{fig: restoring stress}b and c both illustrate measurements of $\tau$ for four PEGDA concentrations, $c=9$, $19$, $29$, and $39$ \% as a function of UV illumination. We apply a constant UV dosage $\Omega$ in each experimental condition, as calculated from equation \ref{eq: dosage integral}. For each data point shown in Fig. \ref{fig: restoring stress}c, error bars on $\tau$ represent error propagation from the linear fit, following the example in Fig. \ref{fig: restoring stress}a. The behavior of $\tau$ as a function of $\Omega$ is shown in Fig. \ref{fig: restoring stress}b. For drops filled with $c=9$ or $19$ \% PEGDA, $\tau$ remains constant over the entire range of $\Omega$, at $\tau= 232 \pm 80$ Pa.  When $c=29$ and $39$\%, $\tau$ remains below $\sim1000$ Pa at low UV doses.  As $\Omega$ reaches $\sim 3000$ mJ/cm$^2$ and above, illuminating drops with $29$\% PEGDA, $\tau$ increases up to a maximum of $\sim3700$ Pa.  When $c=39$\% PEGDA, the dosage needed to cause an increase in $\tau$ reduces to $\sim1500$ mJ/cm$^2$.  For $c=39$\% PEGDA and $\Omega \gtrsim 1500$ mJ/cm$^2$, $\tau$ increases to a maximum $\sim8200$ Pa.

For emulsion droplets, surface tension can be approximated using $\sigma=\tau a$.  SI Fig. \ref{fig: tau-a vs omega} shows $\tau a$ plotted as a function of $\Omega$, for the same data shown in Fig. \ref{fig: restoring stress}b.  When restoring stress is interpreted as surface tension, $\tau a= 7.3 \pm2.0$ mN/m for $c=9$\% and $\tau a= 5.6\pm 1.6$ mN/m for $c=19$\% PEGDA.  These values are reasonable for surfactant-covered water drops in mineral, with $\chi=10^{-2}$ Span 80.  By comparison, uncured droplets filled with PEGDA exhibit an average surface tension of $\sigma=7.46 \pm 1.39$ mN/m for $c$ between $9\%$ and $39$\% \cite{Shaulsky2025}. This comparison suggests that UV illumination is insufficient to fully cure PEGDA-filled droplets when $c=9$ and $19$\%, and they remain droplets.  However, for drops filled with $c=29$ and $39$\% PEGDA, the increase in $\tau$ with $\Omega$ cannot be reasonably explained by interpreting restoring stress as surface tension.  The maximum values of $\tau$ measured when $c=29$ and $39$\% PEGDA would correspond to $\sigma\sim 80$ and $140$ mN/m, respectively, greater than even the surface tension of bare water droplets in mineral oil, which is 45.6 mN/m.\cite{iqbal2017}  Therefore, when $c=29$ and $39$\% PEGDA, the increase in $\tau$ indicates increased crosslinking and rigidity \added{due to gelation. Like other measures of elastic moduli, the restoring stress also increases with the degree of crosslinking and gelation.} \deleted{This behavior is expected.}\cite{Villone2019, Filatov2017, ChenandAluunmani2022, Jiang2021} 




While $\Omega$ captures the total energy dosage applied to a droplet moving in a fluidic device, $\Omega$ is a combination of two $x$-position dependent variables: UV intensity and exposure. Long exposure time with low UV intensity could yield the same $\Omega$ as short exposure time with high UV intensity. To better understand if $\tau$ is controlled by total dosage or intensity, we plot $\tau$ as a function of intensity in Fig. \ref{fig: restoring stress}c.  Because intensity varies with position, we use a reference position $x=10$ mm, corresponding to the location of the peak in $I_{UV}(x)$ as seen in the UV calibration curve in Fig. \ref{UVschematic3}c.  In Fig. \ref{fig: restoring stress}c, we see that, as $I_{UV_{10}}$ increases up to nearly 1000 mW/cm$^2$, $\tau$ for droplets containing $c=9$ and $19$\% PEGDA remains unchanged.  This behavior is similar to the effect of $\Omega$ on $\tau$ seen in Fig. \ref{fig: restoring stress}b. Interestingly, when $\tau$ is plotted as a function of $I_{UV_{10}}$ instead of $\Omega$, a clearer low-$\tau$ region appears at low intensities for samples containing $c=29$ and $39$\% PEGDA.  Further, there is a much clearer distinction between the behavior at $c=29$ and $39$\% PEGDA.  $\tau$ increases when $I_{UV_{10}}>250$ mW/cm$^2$ for $c=39$\%, but does not increase until $I_{UV_{10}}>800$mW/cm$^2$ for $c=29$\% PEGDA.

The missing link between $\Omega$, shown in Fig. \ref{fig: restoring stress}b, and $I_{UV_{10}}$, shown in Fig. \ref{fig: restoring stress}c, is the exposure time, or residence time of the droplet traveling through the UV illumination region, $t_r$.  Fig. \ref{fig: restoring stress}d explores the effect $t_r$, $I_{UV_{10}}$, and total dose $\Omega$ for $c=39$\% PEGDA. For all fluidics measurements of $c=39$\%, $t_r$ ranges from 3 to 14.8 s, $\Omega$ ranges from 507 to 3475 mJ/cm$^2$, and $I_{UV_{10}}$ ranges from 65 to 952 mW/cm$^2$. The colorbar represents $\tau$; the green line on the colorbar indicates $\tau=230$ Pa, the average $\tau$ for droplets containing $c=9$ and $19$\% PEGDA, seen in Fig. \ref{fig: restoring stress}b.  Interestingly, measurements of $\tau<230$ Pa appear over a range of $t_r$ and $\Omega$, but are mainly limited to $I_{UV_{10}} \lesssim 200$ mW/cm$^2$. Conversely, $\tau$ rises above the threshold for drops only above $I_{UV_{10}} \gtrsim 200$ mW/cm$^2$. Fig. \ref{fig: restoring stress}d demonstrates that although $\Omega$ is a combination of both intensity and exposure time, $I_{UV}$ is likely the limiting variable controlling PEGDA crosslinking. The importance of intensity, rather than dosage matches our understanding of the kinetics of free radical polymerization reactions.  That is, when a photo-initiator is involved in the reaction, it is UV intensity, not dosage, that determines the rate of free radical generation  \cite{krutkramelis2016, dendukuri2008, jariwala2011}. 



\subsection{Local probe of bulk gel crosslinking: gelation time decreases with dosage and concentration}




The above exploration of microparticle crosslinking in flow highlights the complexities of using UV to initiate photo-crosslinking on a moving target. Fluidic results reveal the importance of understanding the relationship between $c$, $\Omega$, $I_{UV}$, $t_r$ and their impact on the crosslinking of the photo-curable polymer PEGDA.  At the same time, investigations of PEGDA crosslinking in bulk, non-emulsified samples are relatively straightforward.  


To assess the time required to crosslink PEGDA, we measure crosslinking kinetics in bulk via the arrest of Brownian tracer particles.  As the PEGDA solution cures due to UV exposure, nanoparticles arrest due to the solidification of the gelled PEGDA.  SI video 1 displays nanoparticles diffusing in 9\% PEGDA solution with a $50 \mu$m scale bar.  As shown in the video, $t_i$ represents the time when the UV light is turned on, which can be identified by a slight increase in image brightness. Nanoparticles then drift to the right before arresting at $t_a$. Although the sample chamber is sealed to reduce drift, unavoidable air pockets remain.  The particle drift is likely motion toward an air pocket on one side of the chamber. As seen in Fig. \ref{fig: kinetics before and after}, striations form in the polymer network in the direction of the light guide after the nanoparticles arrest. The striations are likely from nanoparticles drifting and creating crevices in the hydrogel network.


\begin{figure*}[h!]
\includegraphics[width=\textwidth]{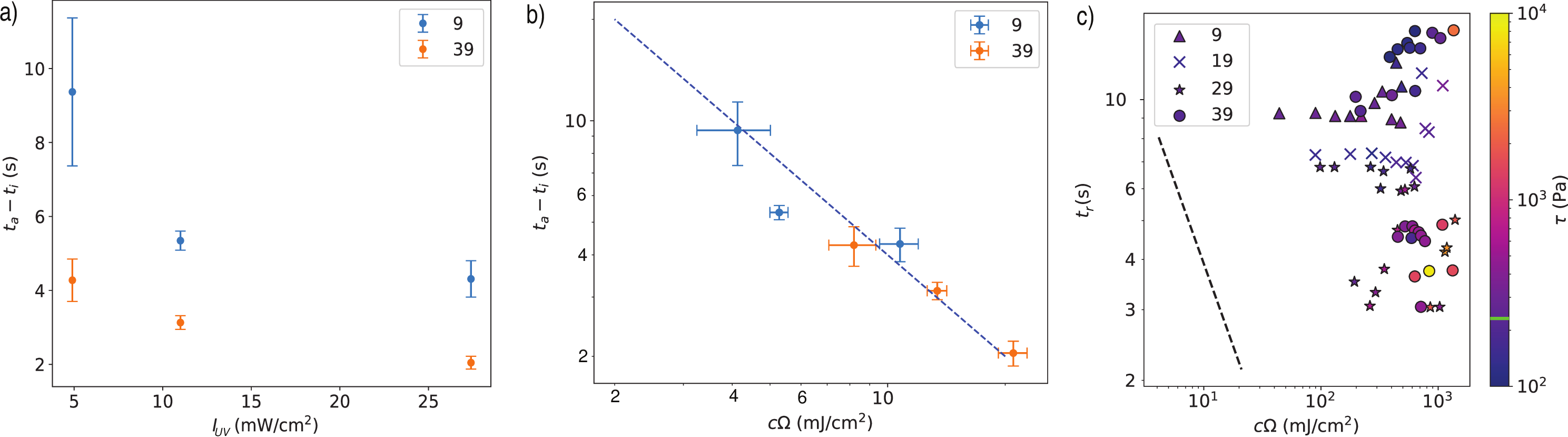}
\caption{Fig \ref{fig: kinetics compilation_1}a: Gelation time in a bulk gel decreases with illumination intensity $I_{UV}$ for bulk gels made of $c=9$ and $39$\% PEGDA. Gelation time is measured as the time between illumination $t_i$ and the arrest of tracer particles, $t_a$.  Error bars represent the standard deviation between three samples. Fig \ref{fig: kinetics compilation_1}b: Gelation time plotted on a log-log axis as a function of $c\Omega$ nearly collapses onto a power law with slope $-1$, as shown by the blue dashed line. Fig \ref{fig: kinetics compilation_1}c: Conditions explored in the fluidic measurements are plotted on log-log axis like those used in b.  Time on the $y$ axis refers to both residence time in the fluidics tests and gelation time in the bulk gels, as indicated by the dashed line.  The color bar indicates measurements of $\tau$ as in Fig. \ref{fig: restoring stress}d.}
\label{fig: kinetics compilation_1}
\end{figure*}

This drift complicates the typical approach of measuring arrest through the mean-squared displacement of the tracer particles.  Instead, we track the instantaneous distance moved by each tracer particle to identify the time of arrest, $t_a$, as the time at which the instantaneous motion decreases to 0.  More details can be found in SI Fig \ref{fig: msd}. Using this method to measure $t_a$, we expose each sample to UV intensities $I_{UV}=4.9$, $11$, and $27.4$ mW/cm$^{2}$. The time to achieve full crosslinking is represented by $t_a-t_i$, and is plotted as a function of $I_{UV}$ in Fig. \ref{fig: kinetics compilation_1}a, for bulk samples composed of $c=9$ and $39$\% PEGDA. The error bars represent the data in triplicate. The smallest UV intensity, 4.9 mW/cm$^{2}$, has the largest error bar: it is more difficult to identify $t_i$ at lower illumination intensities.  As seen in Fig. \ref{fig: kinetics compilation_1}a, the time required to crosslink fully decreases as the UV intensity increases.  Higher PEGDA concentration also leads to a faster crosslinking time.  By noticing that crosslinking time decreases with both $I_{UV}$ and $c$, we normalize $t_a-t_i$ with respect to $c\Omega$, where $\Omega= I_{UV}(t_a-t_i)$. Indeed, as shown in Fig. \ref{fig: kinetics compilation_1}b $(t_a-t_i)$ varies nearly inversely with $c\Omega$.  The blue dashed line on the plot indicates a slope of -1 on the log-log axis.

The power-law relationship between $t_a-t_i$ and $c\Omega$ measured by crosslinking PEGDA in bulk solutions could provide an interesting framework for understanding microparticle crosslinking in flow.  However, there are some important distinctions between the two situations.  In flow, droplets with a given $c$ undergo gelation while flowing through the channel for a particular residence time $t_r$, where they are exposed to UV light with a total dosage $\Omega$.  Droplets filled with PEGDA are suspended in mineral oil, and the entire volume of each droplet flows through the UV illumination region. In measurements of bulk macrogel crosslinking, there is no oil.  Also, gelation is accomplished by localized UV illumination on a region of the sample over an exposure time.  

Despite these distinctions, we superimpose the relationship between $t_a-t_i$ and $c\Omega$ obtained from the macrogel measurements onto the experimental conditions measured in flow.  This is illustrated in Fig. \ref{fig: kinetics compilation_1}c, which shows one data point for each fluidic condition measured across all four PEGDA concentrations. For each fluidic condition, $t_r$ is analogous to $t_a-t_i$ obtained from arrest in bulk gels.  Fig. \ref{fig: kinetics compilation_1}c shows that all fluidic conditions lie above the black dashed reference line, indicating that all droplets in fluidic tests are exposed to an amount of UV light which would suffice to crosslink corresponding bulk formulations.  However, only a subset of the fluidic conditions, where $c=29$ and $39$\% PEGDA, lead to the increases in $\tau$ seen in Fig. \ref{fig: restoring stress}a. This comparison between macrogel gelation time and fluidic conditions suggests that an understanding of bulk gelation is not sufficient to describe droplet gelation in flow.

\subsection{Capillary micromechanics reveals morphology of partial gelation in individual droplets}




\begin{figure}[h!]
\includegraphics[width=\columnwidth]{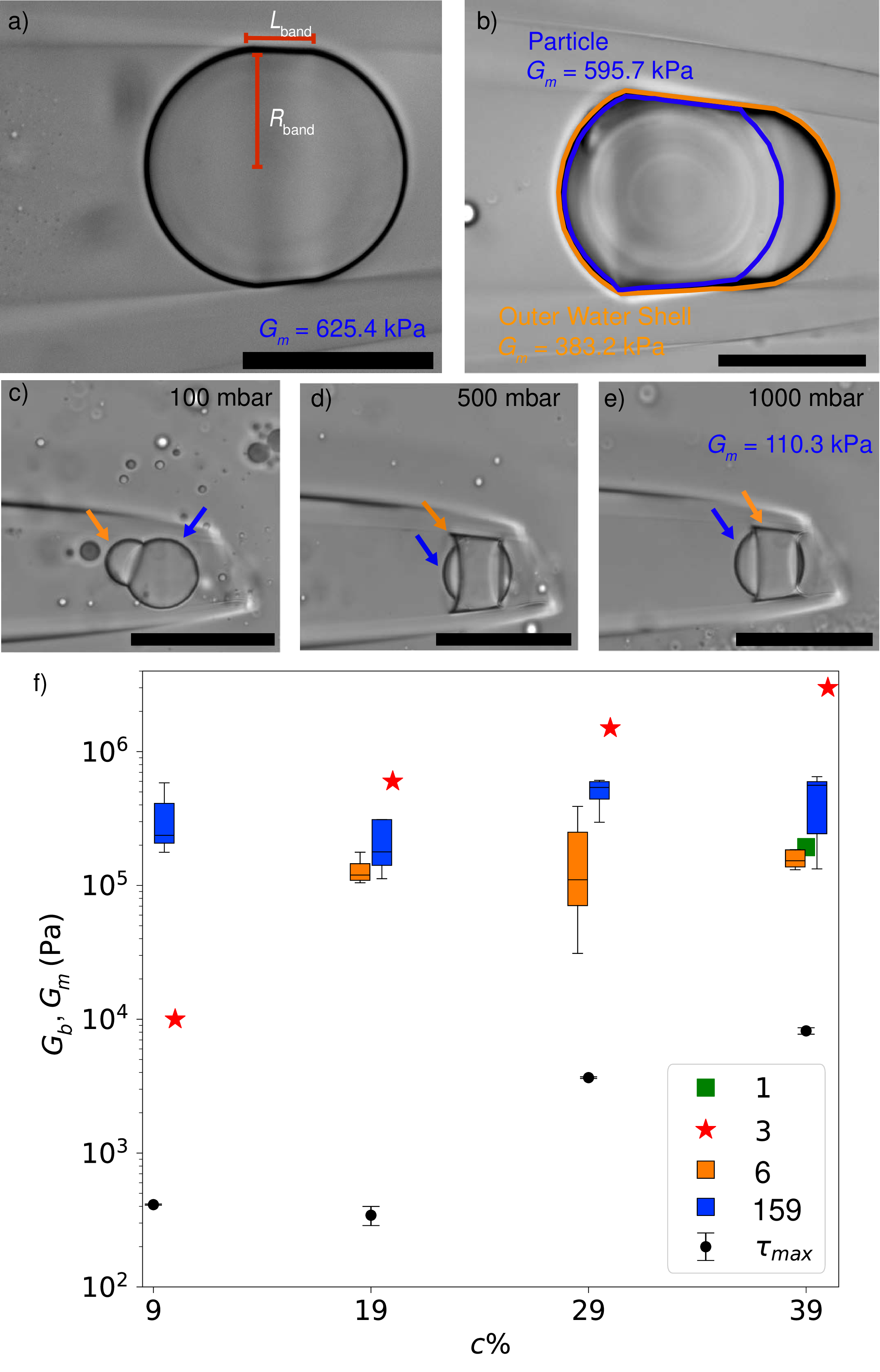}
\caption{Fig.s \ref{fig: capillary micromechanics}a-e show examples of capillary micromechanics measurements.  The examples correspond to: Fig. \ref{fig: capillary micromechanics}a, $c=19\%$ homogeneous gel particle after exposure to 159 J/cm$^2$ with labels for $R_{band}$ and $L_{band}$, Fig. \ref{fig: capillary micromechanics}b, 29\% PEGDA after exposure to 159 J/cm$^2$, Fig. \ref{fig: capillary micromechanics}c-e,  $c=29\%$ PEGDA after exposure to $\Omega=6$ J/cm$^2$ with pressure increasing from left to right, from $100$ to $500$ to $1000$ mbar. In Fig.s \ref{fig: capillary micromechanics}a-e, blue notations refer to gels while orange notations refer to uncured liquid; all scale bars are $50\mu$m.  Fig. \ref{fig: capillary micromechanics}f shows a compilation of measurements of $G_m$ of gels in the capillary micromechanics compared to measurements of $\tau$ from fluidics and $G_b$ from bulk rheology on PEGDA gels of the same chemistry. \cite{Anselmo2015}}
\label{fig: capillary micromechanics}
\end{figure}


The exposure times encountered in fluidics fall well beyond the time required to initiate crosslinking in the localized macrogel kinetic experiments. Therefore, we may expect some degree of crosslinking in the fluidics results for all conditions.  However, the fluidic experiments also yield measurements of $\tau$ consistent with the presence of droplets. To further investigate, we perform independent measurements of crosslinking in individual bulk emulsion droplets gelling in oil across a range of UV dosages and intensities.

We measure shear modulus, $G_m$, by applying increments of pressure to a microparticle immobilized near the exit of a tapered capillary \cite{wyss2010}. Increasing the applied pressure moves the microparticle slightly further toward the capillary exit, causing it to compress and elongate. The relationship between stress and strain provides the modulus.  We measure differential stress as a function of the difference between longitudinal($\epsilon_L$) and radial($\epsilon_R$) strain. $\epsilon_R$ and $\epsilon_L$ depend on the radius ($R_{band}$) and length ($L_{band}$) of the particle touching the capillary walls, as shown in Fig. \ref{fig: capillary micromechanics}a. Due to the taper angle $\beta$, the actual pressure applied on a trapped particle is not $p$, but rather
\begin{equation}
    p_{wall} = \frac{R_{band}}{2L_{band}sin(\beta)}p
    \end{equation}
\noindent Then, $G_m$ is obtained from the slope of $(p_{wall}-p)/2$ versus $\epsilon_R-\epsilon_L$. 

For capillary micromechanic experiments, we make batch microparticles by exposing an emulsion to three different values of $I_{UV}$ and $t$ to yield a wide range of $\Omega$, as seen in table \ref{tab: cap micromech ItD}. The range of $\Omega$ and $I_{UV}$ falls both within the range and beyond that of the fluidic experimental conditions. 


We observe three distinct particle architectures as we push particles through a pulled capillary. The first (Fig. \ref{fig: capillary micromechanics}a) shows an example of a deformed homogeneous particle.  In this image, a droplet with $c=19\%$ has been exposed to $\Omega=159$ J/cm$^2$, resulting in a measurement of $G_m=625.4$ kPa. Fig. \ref{fig: capillary micromechanics}b shows an inner homogeneous particle, outlined in blue, surrounded by what appears to be either a partially cured or uncured outer shell, outlined in orange. The image corresponds to $c=29\%$ and $\Omega=159$ J/cm$^2$. When we measure $G_m$ for both the outer shell and inner particle, we find that $G_m$ for the particle, $\sim600$ kPa, is 1.5 times larger than that of the outer shell, $G_m<400$ kPa. In Fig. \ref{fig: capillary micromechanics}c-e, with $c=29\%$ and $\Omega=6$ J/cm$^2$, a particle, indicated by the blue arrow, is observed along with a droplet, indicated by an orange arrow.  When $p=100$mbar, the droplet adheres to the left or upstream region of the particle (c). As we increase pressure to 500 mbar (d) and 1000 mbar (e), the water or uncured PEGDA drop moves around the particle into a cylindrical structure and is then pushed further down the capillary cavity. In some cases, the uncured liquid is removed entirely from the inner gel material.


Fig. \ref{fig: capillary micromechanics}f presents measurements of $G_m$, using values for cured particles only, disregarding any uncured liquid shells which may be present. The box and whisker plots in Fig. \ref{fig: capillary micromechanics}f represent at least three particles for each condition. The legend labels each set of micromechanics data by the UV dosage used, which we discuss in turn.  At $\Omega=1$ J/cm$^2$, in green, only one data point is obtained, for the highest concentration $c=39$\% formulation.  No other measurements could be obtained.  Capillary micromechanic experiments rely on clogged or fully arrested particles. None of the samples arrest within the capillary for $\Omega=1$ J/cm$^2$ except at $c=39\%$.  Even at pressures as low as 20 mbar, all other compositions deform easily, flowing slowly through and out of the capillary without arresting. Significant deformation and absence of arrest even at very low pressures signify either particles that are extremely soft or droplets. At $c=39\%$ and $\Omega=1$ J/cm$^2$, the green box represents a single particle only; all other measurement attempts resulted in deformation and flow out of the capillary at $p=20$ mbar. The absence of particles at $\Omega=1$ J/cm$^2$ agrees with the fluidic results shown in Fig. \ref{fig: tau-a vs omega}, where almost all measurements are consistent with $\sigma$ of PEGDA-filled drops.  Capillary micromechanics similarly implies that low UV dosage results in droplets for all concentrations. 

At a dosage corresponding to roughly the maximum applied in flow, $\Omega = 6$ J/cm$^2$, in orange, particles immobilize within the capillary for all concentrations except $c=9\%$. At $c=9\%$ and $\Omega = 6$ J/cm$^2$, we observe the same phenomenon described for $\Omega = 1$ J/cm$^2$: drops or soft particles squeeze, deform, and flow through the end of the capillary even at $p=20$ mbar. This result again agrees with fluidics: measurements at $c=9\%$ PEGDA and $\Omega \sim6$ J/cm$^2$ in flow remain consistent with the surface tension of PEGDA-filled droplets. Some of the data obtained at $\Omega = 6$ J/cm$^2$ represent the inner gel portion of a drop that has had its uncured water shell removed. $\Omega=159$ J/cm$^2$, in blue, represents a dosage several orders of magnitude higher than obtained in flow. Particles are captured within the tapered end of the capillary for all PEGDA concentrations, with an average $\langle G_m\rangle \thicksim 400\pm 200$ kPa. This result indicates a dosage and intensity sufficient to fully crosslink droplets into particles.


We compare $G_m$ to two other measurement approaches.  The black dots in Fig. \ref{fig: capillary micromechanics}f show $\tau_{max}$ obtained from fluidics, at dosages below $\Omega\sim 6$ J/cm$^2$.  In fluidics, $\tau$ increases by more than an order of magnitude with increasing $c$, remaining well below $G_m$ measured by capillary micromechanics at $\Omega = 6$ J/cm$^2$.  We include literature values of the bulk gel shear modulus, $G_b$, obtained by oscillatory rheology of bulk macro-sized PEGDA hydrogels, shown as red stars in Fig. \ref{fig: capillary micromechanics}f.\cite{Anselmo2015} The results cited use the same molecular weight PEGDA, photo-initiator, and photo-initiator concentrations as our emulsified PEGDA solutions.  The hydrogels are exposed to $\Omega=3$ J/cm$^2$, corresponding to $I_{UV}=150$ mW/cm$^2$ for $t=20$ s.  This level of UV illumination is within the range of that used in our fluidics measurements. Notably, $G_b$ of the bulk hydrogel increases by two orders of magnitude as $c$ increases from $9$ to $39$\%.

Fig. \ref{fig: capillary micromechanics}f) compares modulus measurements from three very different techniques and on three different types of samples.  The material differences can explain the observations in the comparison.  In bulk gels, modulus intuitively increases with polymer concentration.  In capillary micromechanics, the measurement is of particles prepared from an emulsion.  The presence of the liquid shell seen in Fig. \ref{fig: capillary micromechanics}b-e) complicates the measurement.  The somewhat invasive nature of the capillary micromechanics technique alters the material in that it can remove the water shell.  This removal of material may cause measurements of $G_m$ to be artificially high at low PEGDA concentrations.  In contrast, the gentle shear flows applied in fluidics do not damage the material, and any water shell is allowed to remain intact.  As such, the measurements of $\tau_{max}$ represent more holistic measurements of the entire suspended material, including the effect of any water shell around an internally crosslinked particle.



\subsection{Gelation dynamics in hanging drops confirm the presence of an uncured water shell}


\begin{figure*}[h!]
\includegraphics[width=\textwidth]{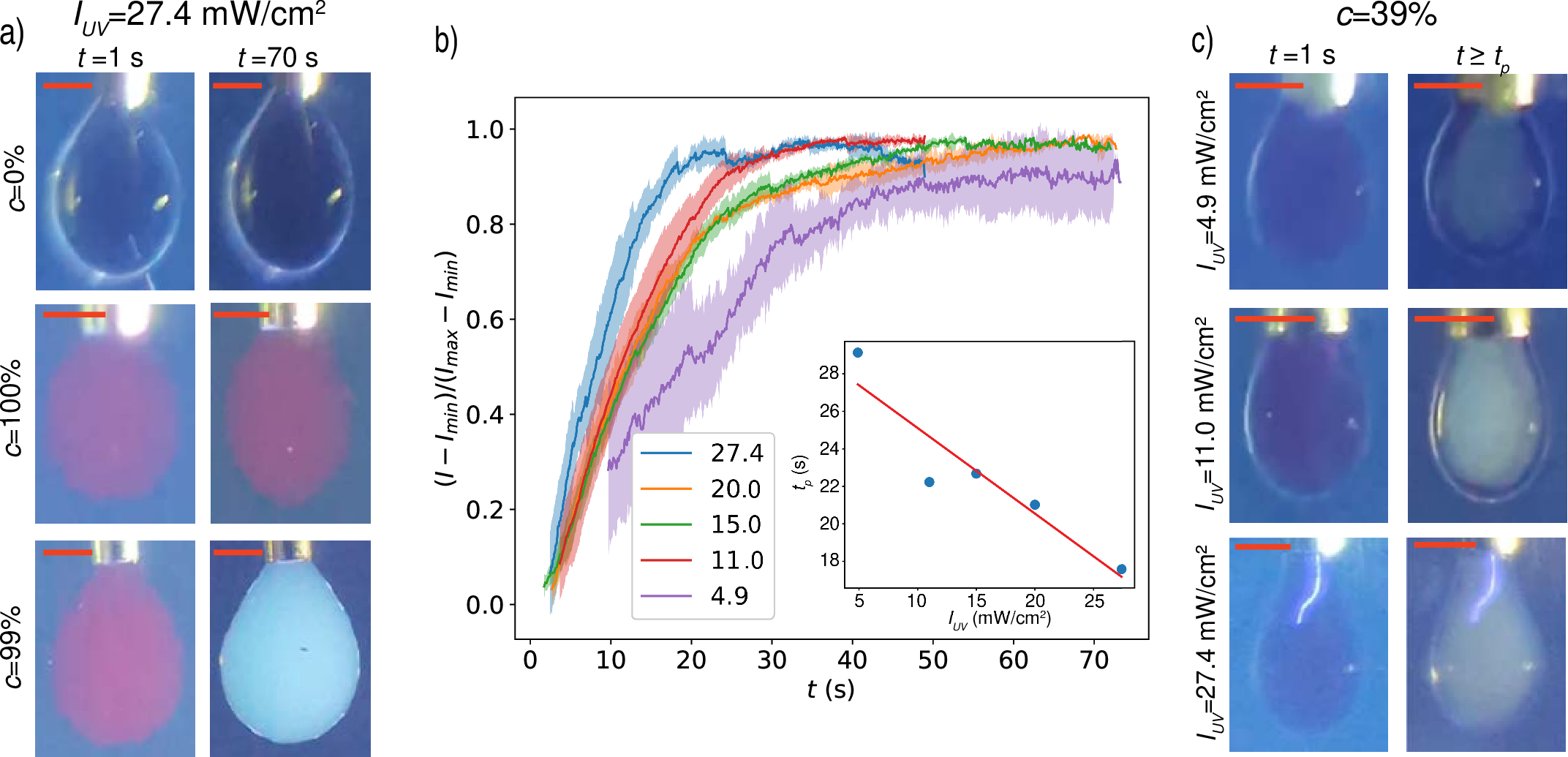}
\caption{Fig \ref{fig: pendant drop images}a: Controls of pendant droplet set-up exposed to 27.4 mW/cm$^2$ with images of exposure at 1 s and 70 s. The first row depicts a sample of water with no PEGDA. The second row depicts $c=100\%$ PEGDA with no photoinitiator. The last row displays $c=99\%$ PEGDA with 1\% photoinitiator. \deleted{Red error bars represent 0.5 mm.} Fig. \ref{fig: pendant drop images}b: Normalized grayscale image intensity versus exposure time, $t$, for $c=39\%$ PEGDA and five UV intensities. Shaded regions represent the deviation of three samples per UV exposure. The inset is the time for the normalized image intensity to plateau ($t_p$) as a function of $I_{UV}$. Fig. \ref{fig: pendant drop images}c: Pendant drop images of $c=39\%$ PEGDA at $t=1$ s and $t\geq t_p$ for three different UV intensities, intensity increasing from the top row to the bottom row. \added{The red scale} \deleted{Red error} bars \added{in Fig \ref{fig: pendant drop images}a and Fig \ref{fig: pendant drop images}c} represent 0.5 mm.}
\label{fig: pendant drop images}
\end{figure*}



Capillary micromechanic measurements reveal interesting features of the particle and droplet morphologies on an individual basis. To confirm the presence of a liquid shell in a less invasive environment, we use a pendant drop or hanging droplet setup to visualize crosslinking dynamics for droplets suspended in oil. 



We add purple dye to PEGDA droplets to enhance image contrast and visualize crosslinking. We first measure several control systems to determine the effect of UV light on the dye. Fig. \ref{fig: pendant drop images}a shows dyed droplets of water, $c=0\%$, in the top row, pure PEGDA with no photo-initiator, $c=100\%$, in the middle row, and $c=99\%$ PEGDA with 1\% photoinitiator in the bottom row. We expose each sample to 27.4 mW/cm$^2$ and show images both at the very beginning of the exposure time, at $t=1$s in the left column, and after $t=70$s of exposure, in the right column. We avoid images near $t=0$ due to the momentary brightness of the UV light turning on.  With dye dissolved in water only, the droplet is blue and remains blue even after 70 s; it does not change color.  Similarly, with dye dissolved in a solution of 100\% PEGDA, without UV initiator, the droplet is purple and remains purple even after 70 s.  Exposure to UV does not affect the color of the dyed water, in the top row, nor the color of uncrosslinked PEGDA, in the second row.  When dye is added to a drop of 99\% PEGDA, with 1\% UV initiator, the drop initially appears purple.  As UV crosslinks PEGDA, however, the color of the droplet changes from purple to light blue. The fact that crosslinking induces a color change in the drops allows us to determine the degree of PEGDA crosslinking by visual inspection.  Fig. \ref{fig: control pendant drop}a shows a grayscale version of Fig. \ref{fig: pendant drop images}a, to confirm that the intensity of the image is sufficient to track the color change. 





We quantify PEGDA crosslinking by measuring image intensity over time of 39\% PEGDA droplets, with initiator, exposed to five different UV intensities. SI video 2 provides an example video of a droplet with $c=39\%$ PEGDA exposed to 11 mW/cm$^2$ in the pendant drop set up with a needle size of 0.5 mm. In the video, we see a dark background and a dark blue droplet that is then exposed to UV light at $t=0$, as indicated by a momentary brightness in the video. As the droplet is exposed to UV light, the inside region turns a brighter blue while there remains a darker blue outer shell surrounding it. Fig. \ref{fig: pendant drop images}b compiles the normalized intensity, in a cropped region right around the droplets as shown in Fig. \ref{fig: pendant drop images}a, as a function of exposure time, with each data set representing a different applied UV intensity, indicated in the legend. The shaded area of the line accounts for the variation between three separate videos. For each UV intensity, the normalized intensity first increases, then reaches a plateau. We measure the time to reach the plateau, $t_p$, as a proxy for the time required to crosslink the droplet.  We calculate a rolling average intensity over 100 frames, or $\sim1.7$s, and define $t_p$ as the time at which the slope is less than 0.02, indicating constant intensity. The inset of Fig. \ref{fig: pendant drop images}b shows $t_p$ as a function of $I_{UV}$.  The red line of the inset of Fig. \ref{fig: pendant drop images}b is a linear trendline with an $R^2$ value of 0.87. Consistent with localized bulk gel kinetics, $t_p$ decreases as $I_{UV}$ increases, indicating faster crosslinking at higher UV intensities. As gelation completes, the gel clogs the end of the needle.

\added{The hanging droplets are exposed to UV intensities that are up $O(100)$ times less than the maximum intensity $I_{UV_{10}}$ reported in flow tests, in Fig. \ref{fig: restoring stress}c.  However, the total energy dosage $\Omega$ in the pendant drop measurements is comparable to that applied to droplets in flow.  The UV intensities indicated in Fig. \ref{fig: pendant drop images}c, when multiplied by the exposure time of 70s, provide $\Omega$ between 300-2000 mJ/cm$^2$, overlapping with the range shown in Fig. \ref{fig: restoring stress}b.}

The range of UV intensities used in Fig. \ref{fig: pendant drop images} matches that used in the bulk gel crosslinking kinetics measurements.  However, $t_p$ is a factor of 3 to 10 longer than $t_a-t_i$ measured from bulk gel crosslinking. This is likely because, in the bulk gel kinetics measurements, only a portion of the sample is exposed to UV, and localized crosslinking measured in a microscopic field of view.  In the hanging drop, the entire drop is exposed to UV.  Further, the color change of the drops likely responds to the completion of the gelation reaction throughout the entire volume of the drop.  The difference in UV illumination is shown in SI Fig. \ref{fig: SIschematic}.



Fig. \ref{fig: pendant drop images}c provides example images of droplets exposed to three UV intensities: 4.9, 11.0, and 27.4 mW/cm$^2$. The left column of Fig. \ref{fig: pendant drop images}c shows an image of the droplet before gelation, after only 1 second of UV exposure. The column on the right of Fig. \ref{fig: pendant drop images}c shows the droplets at a time when $t\ge t_p$. After curing at ($I_{UV}=4.9$ mW/cm$^2$), at $t\ge t_p$, the inside of the droplet is light blue with a surrounding darker blue shell, as seen in the top row. This effect persists with $I_{UV}=11.0$ mW/cm$^2$ and $t\ge t_p$, as seen in the middle row: a light blue center is surrounded by a darker blue outer shell. However, the outer dark blue region is less pronounced with the increase in UV intensity.  At the highest UV intensity, in the bottom row, the droplet is uniformly light blue at $t\ge t_p$, and there is little to no blue shell. 

Fig. \ref{fig: control pendant drop}b shows the grayscale images of Fig. \ref{fig: pendant drop images}c. Additionally, in Fig. \ref{fig: control pendant drop}b, we include an example droplet of $c=9\%$ exposed to $I_{UV}=27.4$ mW/cm$^2$. At $c=9\%$, the image intensity does not exhibit a plateau within 80 seconds, which is the time threshold of the UV light, and at intensities lower than 27.4 mW/cm$^2$. In the one example droplet exhibiting a plateau in intensity, at $c=9\%$ and $I_{UV}$=27.4 mW/cm$^2$, the droplet changes colors from dark blue at $t=1$ s to light blue $t\ge t_p$ s. At $t\ge t_p$ s, there is no outer dark blue shell, only a homogeneous light blue color. 

The color change to light blue indicates PEGDA crosslinking, while the remaining dark blue indicates uncured water. We deduce that the inner light blue region observed in Fig. \ref{fig: pendant drop images}c corresponds to regions of crosslinked PEGDA. The outer shell of the particles that remains a darker blue is likely uncrosslinked PEGDA or water. As UV intensity increases, the outer shell decreases in thickness. Microscopy measurements of tracer particle diffusion corroborates gelation in the internal region of the droplet.  Particles in the outer region continue diffusing after those in the center arrest due to gelation after UV exposure, as seen in Fig. \ref{fig: drop diffusion}. 



The presence of an outer shell of unpolymerized material may be due to oxygen inhibition of the free radical polymerization.\cite{decker1985,obrien2006,ligon2014} Once UV splits the photoinitiator into free radicals, they can either proceed to polymerize PEGDA monomers, or they can be consumed by oxygen.  When PEGDA droplets are exposed to UV light, any oxygen present in the liquid surrounding the droplet diffuses into the droplet from all sides, competing with the polymerization reaction in the outer region of the droplet.  The inhibition reaction is faster than polymerization; hence, polymerization proceeds only in regions of the droplet where oxygen has been entirely consumed.  Because UV light breaks the photoinitator into radicals throughout the entire volume of the drop, polymerization can proceed in the center.  This phenomenon has been observed in PEGDA droplets fabricated in fluidic flows and crosslinked by UV light while flowing. The unpolymerized shell of microfluidic-made PEGDA particles can be reduced or eliminated by creating a nitrogen jacket around the fluidic device.\cite{krutkramelis2016} \deleted{At the same time, oxygen solubility in mineral oil is very low.} Another possible explanation for this uncured liquid around an inner gel is the expulsion of water from the gelled regions.  PEGDA contains acrylate groups, which, when assembled together, can exhibit hydrophobic behavior \cite{barney2022strength}.

\added{To assess the possibility that oxygen quenches PEGDA crosslinking, we purge the aqueous phase and/or the continuous mineral oil phase with nitrogen for two minutes before exposing the pendant drop system to UV light. As seen in Fig. \ref{fig: purge}, purging mineral oil with nitrogen is sufficient to remove the aqueous liquid layer surrounding the cured gel particle. This suggests the fluid layer surrounding the inner gel region, seen in Fig. \ref{fig: pendant drop images}c, can be explained by the diffusion of oxygen into the droplet and its subsequent consumption of free radicals.}

\subsection{Manipulation of the uncured liquid shells in flow generates a variety of structures}




Both capillary micromechanics and pendant drop measurements reveal the presence of a water shell around an internal region that undergoes gelation. Further, the pendant drop measurements suggest the size of this layer decreases with increasing UV intensity.  These two observations, combined, may provide insight into the fluidics approach discussed in Fig. \ref{fig: restoring stress}.  That is, the presence of an uncured water shell around a gelling particle might contribute to the low values of restoring stress measured in flow.   

In addition to the non-invasive fluidic approach described above, we also run flow tests in a device geometry that includes tight constrictions along the flow path.  This alternate device design, shown in Fig. \ref{UVschematic3}b, allows manipulation and alteration of the PEGDA droplets and particles while they flow through it.


\begin{table}[h!]
\resizebox{\columnwidth}{!}{\begin{tabular}{cccccccc}
\centering
\textbf{State} & \textbf{$\Omega$} &\textbf{$I_{UV_{11}}$} & \textbf{c} & \textbf{$\sigma$} & \textbf{$\Gamma$} & \textbf{Fig.} & \textbf{Vid.} \\

\textbf{} & \textbf{(mJ/cm$^2$)} & \text{mW/cm$^2$} & \textbf{($\%$)} & \textbf{ (mN/m)} & \textbf{(ms)} & \textbf{} & \textbf{} \\
\hline

Uncured               & 1000               & 171.9                   & 9\%               & 3.56                            & -                      & Y                     &                      \\
Uncured               & 2700          &  504.2                            & 9\%               & 3.9                             & -                      &                       & Y                    \\
Uncured               & 5000              & 339.9                          & 9\%               & 6.7                             & -                      &                       &                      \\

Uncured       & 6000                & 339.9                        & 9\%               & -                               & -                      & Y                     &                     \\

Partial        & 6000                & 924.3                        & 9\%               & -                               & -                      & Y                     &                      \\
Partial        & 6200                 &   924.3                    & 9\%               & -                               & -                      & Y                     & Y                    \\
Full            & 11000                   & 924.3                    & 9\%               & -                               & 102                 & Y                     & Y                    \\
\hline
Uncured               & 1500        &125.4                                & 19\%              & 3.1                             & -                      &                       &                      \\
Uncured               & 2400               &313.6                         & 19\%              & 4.7                             & -                      & Y                     &                      \\

Partial        & 3000                    &313.6                    & 19\%              & -                               & -                      & Y                     & Y                    \\
Partial           & 3300                      &254.7                  & 19\%              & -                               & 4.9                  & Y                     &                      \\
Partial           & 3300                  &254.7                      & 19\%              & -                               & -                      &                       & Y                    \\

Partial           & 3500                  &250.8                      & 19\%              & -                               & -                      & Y                     &                      \\
Partial           & 4100                       &672.2                 & 19\%              & -                               & -                      & Y                     &                      \\

Full           & 4500              &924.3                          & 19\%              & -                               & -                      & Y                     &                      \\
Full           & 4600           &924.3                             & 19\%              & -                               & 10.2                & Y                     & Y                   
\end{tabular}}
\caption{Experimental formulation conditions for fluidic tests performed in the device design shown in Fig. 1b.}
\label{tab: transient table}
\end{table}

\begin{figure*}[h!]
\includegraphics[width=\textwidth]{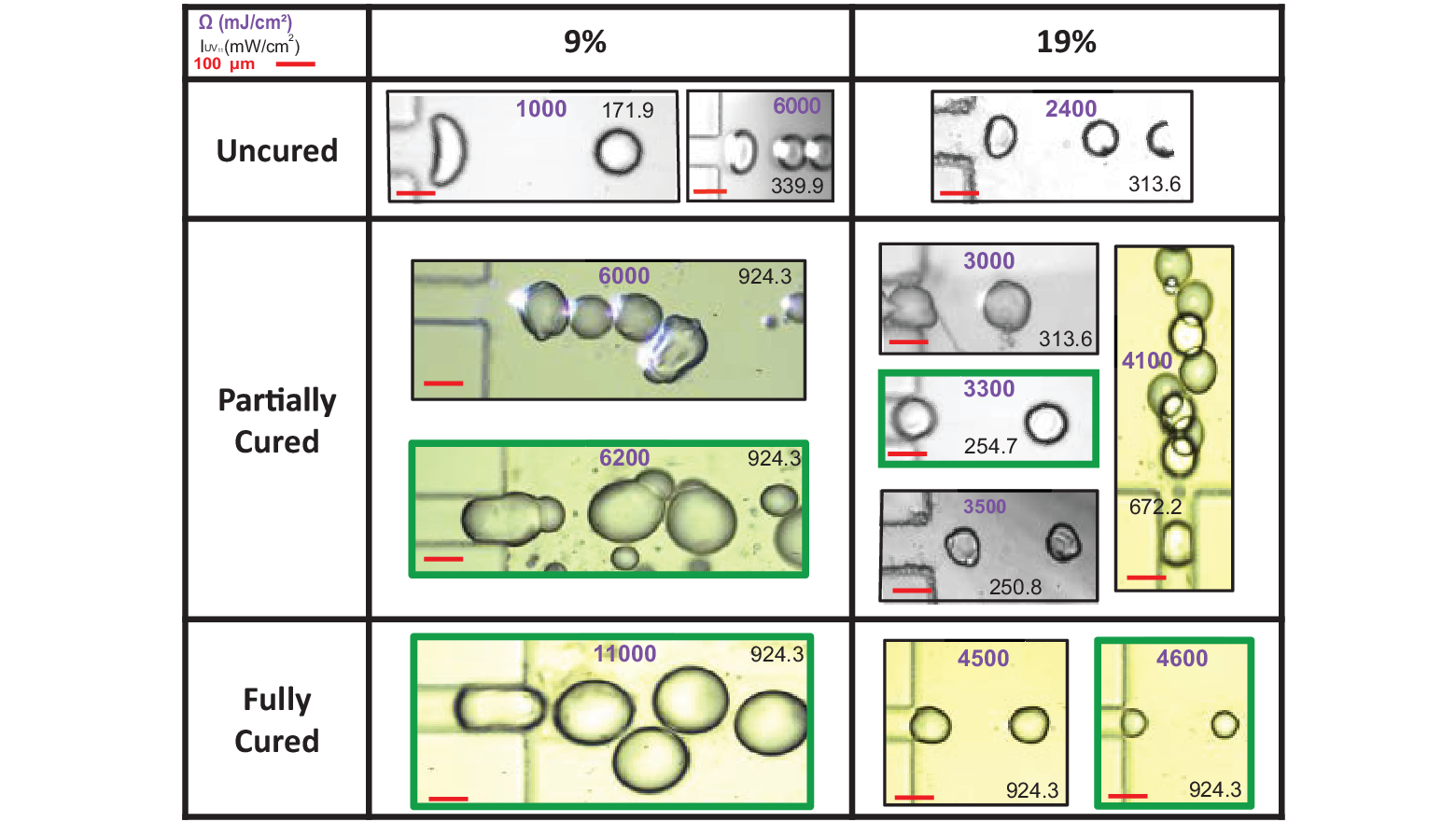}
\caption{Each panel presents microscopy images of uncured, partially cured, and fully cured droplets.  Conditions for each panel are fully described in Table \ref{tab: transient table}. The images outlined by green boxes are accompanied by SI video footage.}
\label{ParticleImageTable}
\end{figure*}

Table \ref{tab: transient table} indicates the experimental conditions explored, using compositions of $c=9$ and $19$\% PEGDA. We focus on behavior after the first constriction encountered in the flow direction, $100\mu$m in width. A slightly different UV light guide angle applies and $I_{UV}$ is maximum at $x=11$ mm rather than 10 mm. For this reason, the Table refers to intensity as $I_{UV_{11}}$.  Table \ref{tab: transient table} indicates which conditions are shown in the microscopy images in Fig. \ref{ParticleImageTable}, and which are illustrated by video clips in the SI.   

The first column in Table \ref{tab: transient table} indicates the `State' of the material observed in microscopy.  A representative selection of these behaviors is shown in Fig. \ref{ParticleImageTable}.  The images in Fig. \ref{ParticleImageTable} are labeled in purple with the UV dosage, in black with the UV intensity, and with red scale bars of $100\mu$m.  The `Uncured' state refers to droplets.  Upon exiting the constriction, these elongate in the direction transverse to the flow before relaxing back to spheres.  This is seen at sufficiently low UV illumination at both $c=9$ and $19$\%, as in the top row of Fig. \ref{ParticleImageTable}.  At $c=9$\%, this behavior persists up to $\Omega\sim6000$ mJ/cm$^2$.  As $c$ increases to $19$\%, droplet behavior persists up to $\Omega\sim6000$ mJ/cm$^2$. 

SI video 3 shows the dynamics of this relaxation behavior of droplets exiting a constriction.  Coupled analysis of deformation relaxation dynamics with the details of the extensional flow can be used to measure surface tension. \cite{Hudson2005, chen.dutcher.2020, Shaulsky2025}  We provide details of the analysis in the SI, with the final measurements of surface tension $\sigma$ indicated in Table \ref{tab: transient table}, for all experimental conditions that lead to uncured droplets.  All measurements of $\sigma$ obtained from droplet relaxation dynamics are comparable to the values of $\tau a$ at $c=9$ and $19$\% in Fig. 2c.  


Increased UV illumination results in partial curing, illustrated in 
the middle row of Fig. \ref{ParticleImageTable}.  At $c=9$\%, this transition to partial curing happens at $\Omega\sim6000$ mJ/cm$^2$.  The UV threshold decreases as $c$ increases to $19$\%, to $\Omega\sim3000$ mJ/cm$^2$.  

At $c=9$\%, the image obtained at $\Omega=6000$ mJ/cm$^2$ shows a variety of dimpled or wrinkled shapes, which may result from the curing process persisting in flow through the constriction, during which transit time the material is deformed and elongated.  However, the example image at $\Omega=6200$ mJ/cm$^2$ clearly shows the presence of smaller water drops attached to primary gel particles. 
Much like in the capillary micromechanics technique, flow through the tight constriction can push the uncured liquid to the front of the particle and then ultimately wick it away.  For this reason, three of the four particles seen in Fig. \ref{ParticleImageTable}, with $c=9$\% and $\Omega=6200$ mJ/cm$^2$ have asymmetric shapes, with the water drop remaining attached to the gel on one side of the particle.  In the fourth particle, at the right of the image, the water shell has been completely wicked away from the droplet.  This process can be seen in the SI video of flow through and at the exit of the constriction. 

At 19\% PEGDA, many particles in the `partial curing' regime are seen with several beads of uncured water on the outside of gel particles. When two of these water beads are present, they are often arranged in a shape somewhat similar to that of triatomic bent molecules, like water.  This can be seen in the example with $c=19$\% PEGDA exposed to $\Omega=3500$ mJ/cm$^2$.  In other cases, there is evidence that these small beads of water have been shed from the particles, taking the form of small satellite drops like those seen in the example with $\Omega=4100$ mJ/cm$^2$. Interestingly, at $\Omega=3300$ mJ/cm$^2$, the partially cured particles maintain a nearly spherical shape, but an inner rim can be seen on the inside of the outermost dark contour.




At sufficiently high UV illumination, PEGDA gelation results in particles, referred to in Table \ref{tab: transient table} as being in the ``Full'' curing state.  We identify the fully cured state by investigating deformation relaxation behavior at the exit of a constriction.  In contrast to the behavior of droplets, which first elongate transverse to the flow direction before relaxing, gelled particles exhibit positive values of $D$ as they exit the constriction. This can be seen in Fig. \ref{ParticleImageTable}, with $c=9$\% and $\Omega=11000$ mJ/cm$^2$: slight elongation remains in the flow direction in all materials that have exited the constriction.  The dynamic response of this condition can be seen in the SI video.  In this curing state, $D(t)$ relaxes exponentially to $D\rightarrow0$, and can be fit to $D=\exp(-t/\Gamma)$.  An example is shown in SI Fig. \ref{fig: extension flow SI}c and d.  As seen in Table \ref{tab: transient table}, the condition with $c=9$\%, $\Omega=11000$ mJ/cm$^2$ and $I=924$ mW/cm$^2$ results in this behavior, with $\Gamma=102$ms.  As $c$ increases to $19$\%, a lower dosage, $\Omega=4600$ mJ/cm$^2$ and $I=924$ mW/cm$^2$ results in this same type of exponential relaxation, with $\Gamma=10.2$ms.





Comparison of results obtained at $c=9$ and $19$\% PEGDA corroborates previous results that lower degrees of UV illumination are required to gel systems with higher PEGDA content.  Interestingly, as seen in Table \ref{tab: transient table}, $\Omega=6000$ mJ/cm$^2$ can lead to either droplets, at $I_{UV_{11}}=340$ mW/cm$^2$, or partial curing, at $I_{UV_{11}}=924$ mW/cm$^2$.  This observation matches that discussed in the steady flow tests discussed in Fig. \ref{fig: restoring stress}d: it is high $I_{UV_{11}}$, and not only high $\Omega$, which is required for gelation in flow.

\section{Conclusion}
The suite of experimental techniques and observations discussed here reveals the subtleties of gelation and material property evolution while PEGDA polymer gelation is accomplished in flow, using UV illumination coupled with droplet-based fluidics.  While the fluidic approach reliably generates emulsion drops of monodisperse sizes, achieving controlled uniformity of gelation and mechanical properties requires a careful experimental approach.


In flow, we measure steady deformation of PEGDA-filled droplets while they gel, and use restoring stress as a holistic metric of the mechanical response.  We expect restoring stress to increase as PEGDA droplets become solid microgels. Interestingly, we find that lower concentrations of PEGDA do not exhibit an increase in restoring stress, even at high UV doses up to 6000 mJ/cm$^2$. 

While the low values of restoring stress suggest that low concentrations of PEGDA may not gel properly in flow, orthogonal measurement techniques suggest otherwise.  The kinetics of crosslinking in bulk gels suggest that residence times in fluidic flows are sufficiently long to enable crosslinking.  Interestingly, measurements of gelled microscale and meso-scale droplets both reveal the presence of an outer, uncured water layer around an inner gel.  The presence of this uncured liquid shell could explain why the values of restoring stress measured in flow are equivalent to the surface tension stress of surfactant-covered drops.  At the same time, we find we can use this uncured shell as a lever by which to control the structure and morphology of the PEGDA in its partial curing state.


The above work highlights the importance of understanding the detailed coupling of reaction dynamics, structure, and mechanics for all types of microfluidic particle fabrication.  Despite the fact that most of these materials are made and then used with specific applications in mind, the richness of gelation phenomena in flow is worth exploring and describing.  Further, the ability to make and manipulate uncured water shells around gelled material could facilitate further in-line or downstream interfacial chemistry on gel surfaces in flow.

\section{Supplementary Information}
The Supplementary Information includes: symbols and their  definitions, organized by the order of their appearance in the manuscript; additional schematics showing the geometry of UV illumination; measurements of $\tau a$ as a function of $\Omega$ from fluidics tests; additional descriptions of the bulk gelation kinetics measurements, including a video; a table of experimental conditions investigated through capillary micromechanics; additional description and corroboration of the pendant drop gelation measurements, including a video; measurements of relaxation dynamics associated with Fig. \ref{ParticleImageTable}; a further visualization of steady curing drops in fluidics. 


\section{Conflicts of interest}
The authors have no conflicts of interest at this time. Some
details of this work have been filed as PCT/US23/30252 on
August 15, 2023, and published on February 22, 2024; publi-
cation number WO 2024/039662

\section{Acknowledgments}
We gratefully acknowledge funding from the Northeastern University Center for Research Innovation CRI Spark Fund, for SM, and from the Undergraduate Research and Fellowships Office, for AJP.  \added{We appreciate helpful conversations with Megan Valentine and Petia Vlahovska.} The authors thank Charles Wallace for developing early versions of the code used to track and analyze particles exiting a constriction, and Allison Dennis for the use of her lab's UV-vis spectrophotometer.  



\balance


\bibliography{rsc-articletemplate-softmatter} 
\bibliographystyle{rsc}
\newpage
\clearpage
\input{supplemental}

\end{document}

%% file: supplemental.tex
\setcounter{figure}{0}
\renewcommand{\thefigure}{S\arabic{figure}}

\setcounter{table}{0}
\renewcommand{\thetable}{S\arabic{table}}
\raggedbottom
\onecolumn
\allsectionsfont{\centering}
\pagenumbering{gobble}

\section{Supplementary Information for \\ Subtleties of UV-crosslinking in microfluidic particle fabrication: \\ UV dosage and intensity matter}


\subsection{Symbols}
To establish a standardized nomenclature, we report symbols and their definition in table \ref{table:symbol}, organized by the order of their appearance in the manuscript.
\begin{table}[h!]
\centering
\begin{tabular}{ll}
\hline
Symbol                         & Definition                                           \\ \hline
$c$                            &  Concentration by volume
\\
$\mu$                        & Viscosity
\\
$\chi$                         & Mole fraction                                      \\
$I_{UV}$                       & UV light intensity                                 \\
$\theta$                       & Light guide angle                                  \\
$x$                         & Coordinate along fluidic flow path                   \\
$x, Z$                         & Light guide height                    
\\
$X_1, Z_1$                     & Transverse and longitudinal light guide distance   \\
$R$                            & Distance from light to droplet position        \\
$I_{UV0}$                      & Initial UV light intensity                         \\
$N$                            & Knob number                                       \\
$t_r$                            & Exposure or residence time  
\\
$\Omega$                       & UV dosage                                           \\
$v$                            & Velocity                                           \\
$\dot{\gamma}$                 & Shear rate                                         \\

$H$                            & Channel height                                     \\
$D$                            & Taylor deformation paramter                        \\
$L, B$                         & Moments of inertia in the $x$ and $y$-direction        \\
$\dot{\varepsilon}$            & Extension rate                                     \\
$\Gamma$                       & Relaxation time \\
$p_{wall}$                     & Pressure applied from wall to particle             \\
$p$                            & Applied pump pressure                              \\
$V$                            & Particle volume                                    \\
$R_{band}, L_{band}$           & Radius and length of particle in contact with wall \\
$\beta$                        & Capillary taper angle                              \\
$\varepsilon_R, \varepsilon_L$ & Strain deformations for $R_{band}$ and $L_{band}$  \\
   $\langle s^2 \rangle$                        & Mean-squared displacement                          \\
$t_d$                          & Delay time                                         \\
$\mathcal{D}$                   & Diffusion coefficient                \\
$I$                            & Image Intensity \\
Ca                            & Capillary number 
\\

$\lambda$                     & Viscosity ratio   \\

$\tau$                        & Restoring stress
\\
$a$                           & Droplet radius
\\
$\sigma$                      & Surface tension
\\
$G$                           & Shear modulus
\\
$K$                          & Compressive modulus
\\
$t_i, t_a$                    & Illumination and arrest time
\\
$\alpha$                      & Power-law coefficient 
\\
$\Delta s$                     & Instantaneous distance moved
\\

\end{tabular}
\caption{Definitions of symbols and variables (in order of appearance)}
\label{table:symbol}
\end{table}

\subsection{Supplementary schematics of UV illumination in orthogonal measurement methods}

\begin{figure*}[h!]
\centering
\includegraphics[width=0.75\textwidth]{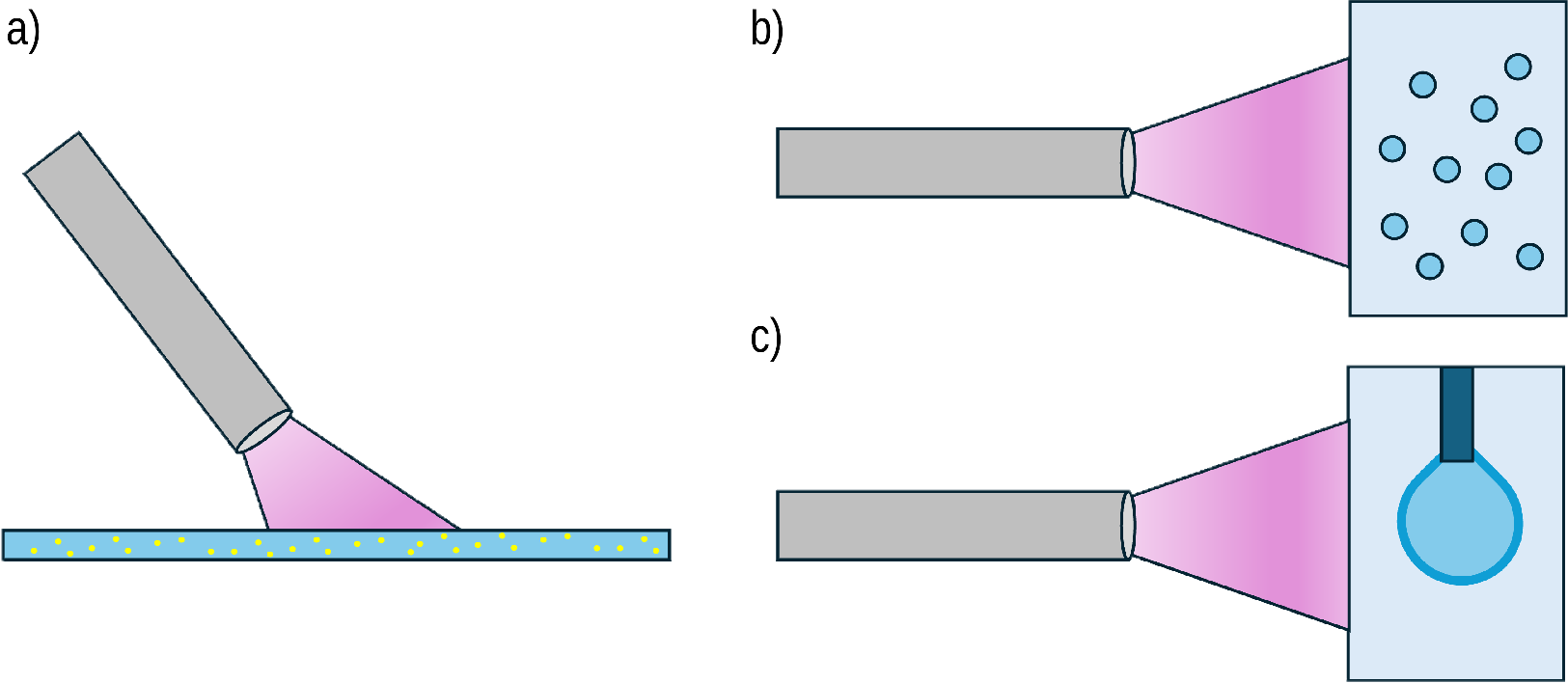}
\caption{Manner of UV illumination for the gelation kinetics measurements in Fig. \ref{fig: SIschematic}a, for the capillary micromechanics measurements in Fig. \ref{fig: SIschematic}b, and for the pendant drop measurements in \ref{fig: SIschematic}c.} 
\label{fig: SIschematic}
\end{figure*}

Fig. \ref{fig: SIschematic} shows the manner of UV illumination for samples prepared for the gelation kinetics measurements in Fig. \ref{fig: SIschematic}a, for the capillary micromechanics measurements in Fig. \ref{fig: SIschematic}b, and for the pendant drop measurements in Fig. \ref{fig: SIschematic}c.  In all three panels the color of the PEGDA solution is blue-green.  In Fig. \ref{fig: SIschematic}a the embedded fluorescent tracer particles are shown in yellow, and the UV illumination is localized to one region of the sample.  In both Fig.s \ref{fig: SIschematic}b and \ref{fig: SIschematic}c the PEGDA solution is prepared in an emulsion form, as droplets in a background of mineral oil with mole fraction $\chi=10^{-2}$ Span 80 surfactant to impart stability.  In both Fig.s \ref{fig: SIschematic}b and \ref{fig: SIschematic}c the UV illumination reaches across the entirety of the drops.

\subsection{PEGDA microparticles crosslink with sufficient PEGDA concentration, UV dosage, and UV illumination}
\begin{figure*}[h!]
\centering
\includegraphics[width=0.5\textwidth]{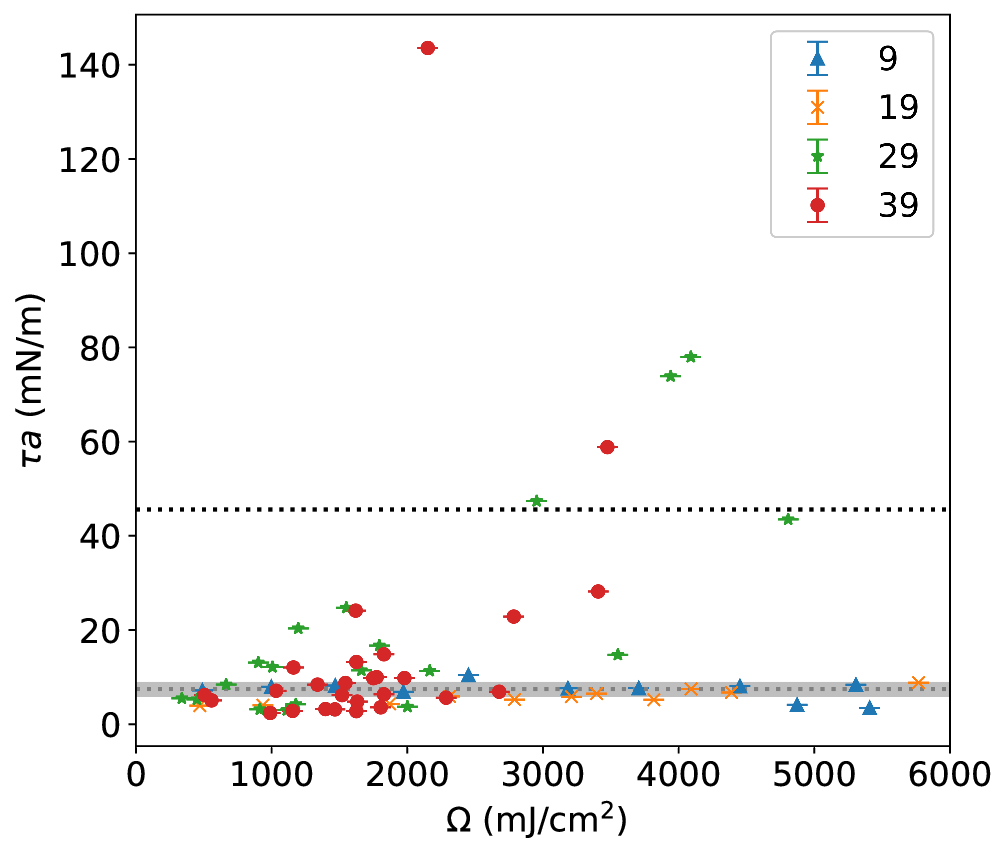}
\caption{Restoring stress multiplied by radius versus UV dosage for four concentrations of PEGDA. Error bars indicate error propagation from the linear fit of $D$ versus $\mu\dot{\gamma}/a$. The black dotted line represents the surface tension of pure water in mineral oil. The gray dotted line represents the average surface tension of 9-39\% PEGDA in mineral oil, with the shaded area representing the standard deviation. \cite{Shaulsky2025}} 
\label{fig: tau-a vs omega}
\end{figure*}

In the fluidic device, we measure $\tau$ for all flowing materials, as a measure of holistic mechanical response. For droplets, the restoring stress $\tau$ is defined as $\sigma/a$. For particles $\tau=G$. Fig. \ref{fig: restoring stress}b shows $\tau$ as a function of $\Omega$. Fig. \ref{fig: tau-a vs omega} shows $\tau a$ plotted as a function of $\Omega$. With $\chi=10^{-2}$ Span 80, pure water droplets exhibit $\sigma<10$ mN/m, as indicated by the gray shaded horizontal line.\cite{Shaulsky2025} By contrast, the surface tension of a water-mineral oil interface without surfactant is $\sigma \sim 45$ mN/m, indicated by the dotted black line.  As seen in Fig. \ref{fig: tau-a vs omega}, we observe multiple measurements of $\tau*a>10$ mN/m, and some of $\tau*a>45$ mN/m, suggesting the presence of materials that are no longer droplets and are likely partially cured or fully cured particles.

\subsection{Localized (bulk) macrogel crosslinking: time required for macrogel crosslinking decreases with dosage and concentration}

\begin{figure*}[h!]
\centering
\includegraphics[width=0.5\textwidth]{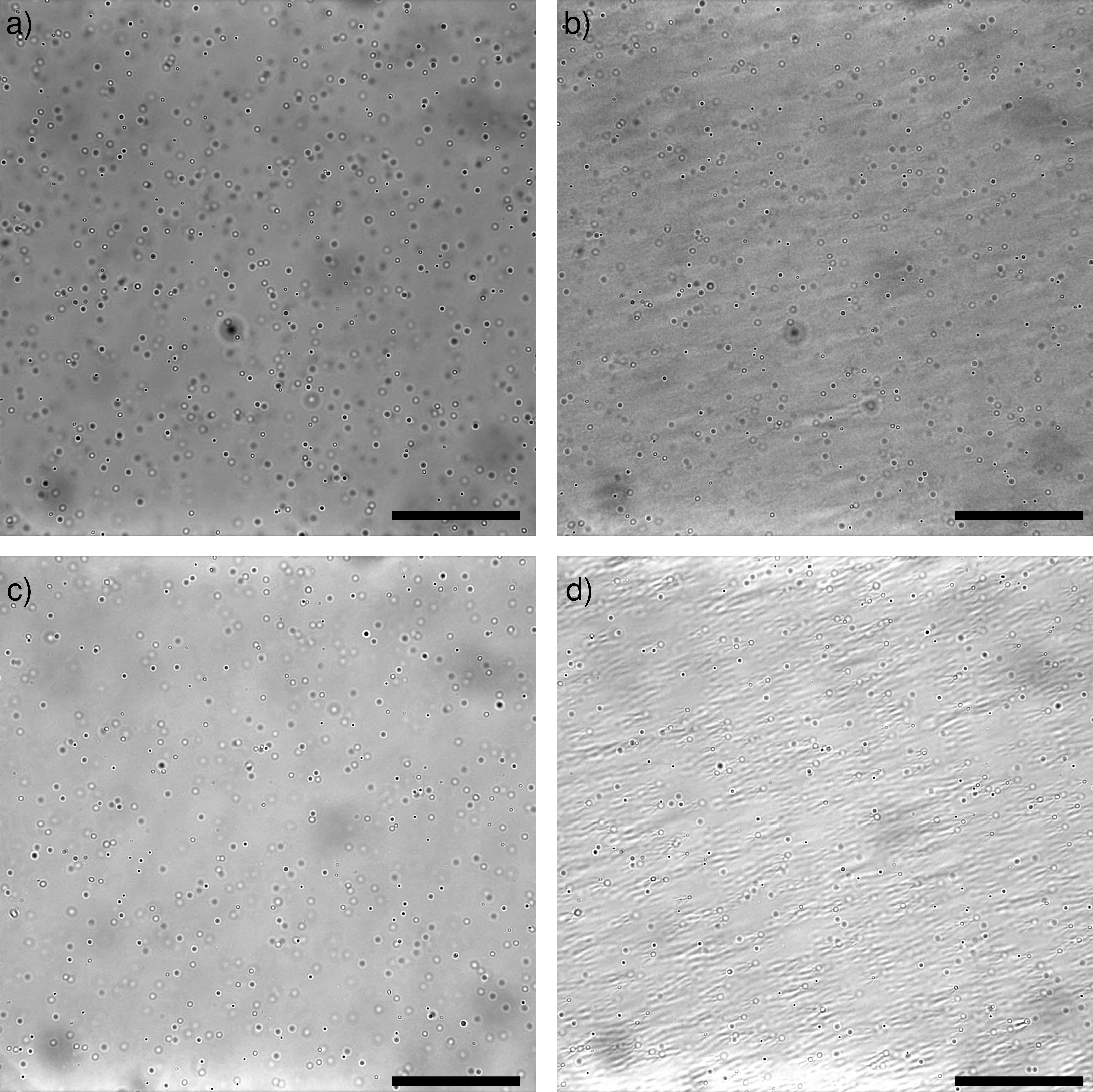}
\caption{Images of 500 nm particles in PEGDA solutions before and after curing at 27.4 mW/cm$^{2}$ for 24.4 seconds. Fig.s S3a and S3b show 9$\%$ PEGDA before and after curing. Fig.s S3c and S3d show 39$\%$ PEGDA before and after curing. The black scale bars are 50 $\mu$m. } 
\label{fig: kinetics before and after}
\end{figure*}

We embed bulk macrogel solutions of 9 and 39\% PEGDA with 500 nm fluorescent particles. In uncured gel solutions, particles passively diffuse due to Brownian motion. When the solutions are exposed to UV light, nanoparticles arrest as the material gels. Fig. \ref{fig: kinetics before and after} shows images of the gel before and after being exposed to 27.4 mW/cm$^2$ with 50 $\mu$m scale bars. We observe striations forming in the gel after exposure to UV light in S3b and S3d, aligned in the direction of the UV light.  The striations are likely caused by nanoparticles drifting in the gel as the gel crosslinks and expels water. SI video 1 shows gelation in a sample of $c=9$\% PEGDA upon exposure to $I_{UV}27.4$ mW/cm$^2$.

\par 

\begin{figure*}[h!]
\centering
\includegraphics[width=0.75\columnwidth]{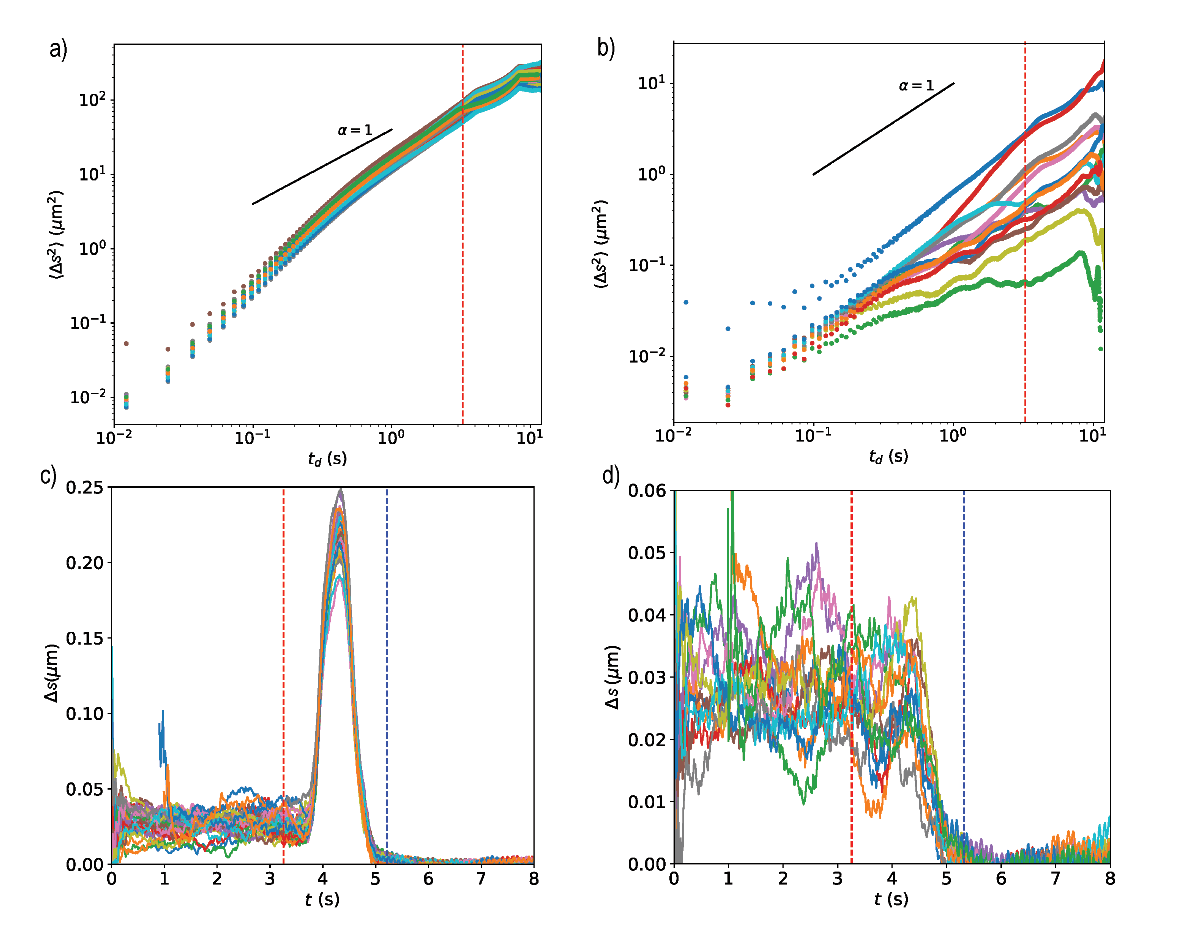}
\caption{Mean-squared displacement for particle tracers in 39$\%$ PEGDA exposed to 27.4 mW/cm$^2$ before drift subtraction, Fig. S4a, and after drift subtraction, Fig. S4b. $\alpha$ is the power-law coefficient of 1. The red dashed lines represent $t_i$. Fig.s S4c and S4d plot the instantaneous distance $\Delta s$ each particle moves as a function of time before, Fig. S4c, and after, Fig. S4d, drift subtraction. The red and blue dashed lines represent $t_i$ and $t_a$, respectively. } 
\label{fig: msd}
\end{figure*}

Fig. \ref{fig: msd} shows our method of assessing gelation time in a bulk sample using nanoparticle tracers. For each solution, at $c=9$ and $39$\% PEGDA, we track particle trajectories before and after illumination by UV light. Typically, the arrest of diffusive motion due to gelation can be measured by tracking particle positions and measuring the mean squared displacement, $\langle \Delta s^2 \rangle$, of particle trajectories as a function of the lag time $t_d$.  Particle diffusion results in power law behavior with slope 1, with arrest signified by a plateau in $\langle \Delta s^2 \rangle(t_d)$.  However, the presence of particle drift complicates this approach. 


Fig. \ref{fig: msd}a and \ref{fig: msd}b show plots of $\langle \Delta s^2 \rangle$ for nanoparticles diffusing in the PEGDA solution and arresting as the gel cures, in a sample of $c=39\%$ PEGDA exposed to 27.4 mW/cm$^2$. The red vertical line indicates $t_i$, or the initial time of UV exposure. $\alpha$ is a power-law coefficient of 1, which indicates Brownian motion. Fig. \ref{fig: msd}a is pre-drift subtraction, and b is post-drift subtraction. The mean-squared displacement Fig.s do not indicate nanoparticle arrest.

Therefore, to assess the moment of arrest, we investigate the instantaneous change in position of the particles between each frame, $\Delta s$, as a function of time, $t$, for each of $\sim30$ particles in suspension. SI Fig. \ref{fig: msd}c shows that $\Delta s$ remains roughly constant during particle diffusion, before UV illumination, between $0.02 \lesssim \Delta s \lesssim 0.04$ $\mu$m of motion from frame to frame.  The vertical red dashed line indicates the illumination time, $t_i$, when the UV light is turned on.  We find $t_i$ by identifying the frame of the video when the sample brightens. The sharp increase in $\Delta s$ shortly after $t_i$ signifies nanoparticle drift.  $\Delta s$ decreases as the drift slows, and then approaches $\Delta s \simeq 0$.  SI Fig. \ref{fig: msd}d shows $\Delta s (t)$ after the constant drift velocity has been subtracted from each particle trajectory. SI Fig. \ref{fig: msd}c confirms that Brownian motion persists for a few seconds after $t_i$. Then the particles slow until they arrest and $\Delta s \simeq 0$. The time of particle arrest, $t_a$ is defined as the time at which the average $\langle \Delta s \rangle$ falls below $0.1$ $\mu$m for all particles over a window of 50 frames. The vertical blue dashed line in SI Fig. \ref{fig: msd}c and S4d indicates $t_a$.

\par 

\subsection{Capillary micromechanics reveals UV illumination is sufficient to gel batch emulsion drops}
PEGDA microparticles are made by exposing an emulsion of PEGDA solution in mineral oil to three different values of $\Omega$ in a manner shown in Fig. \ref{fig: SIschematic}b. For batch microparticles, $\Omega=I_{UV}t$. The corresponding $I_{UV}$ and $t$ for each $\Omega$ is shown in table \ref{tab: cap micromech ItD}.

\begin{table*}[h!]
\centering
\begin{tabular}{|c|c|c|}
\hline

$\Omega$ (J/cm$^2$) & $I_{UV}$ (mW/cm$^2$) & $t$ (s) \\ \hline
1                   & 53.7                 & 18.6   \\ \hline
6                   & 262.6                & 22.7    \\ \hline
159                 & 1994.3               & 79.9    \\ \hline
\end{tabular}
\caption{Intensity and exposure time to yield respective dosages for capillary micromechanics experiments}
\label{tab: cap micromech ItD}
\end{table*}






\subsection{Meso-scale droplet gelation reveals/confirms presence of uncured water \added{layer} \deleted{shell} around cured particle}

\begin{figure*}[h!]
\centering
\includegraphics[width=0.5\textwidth]{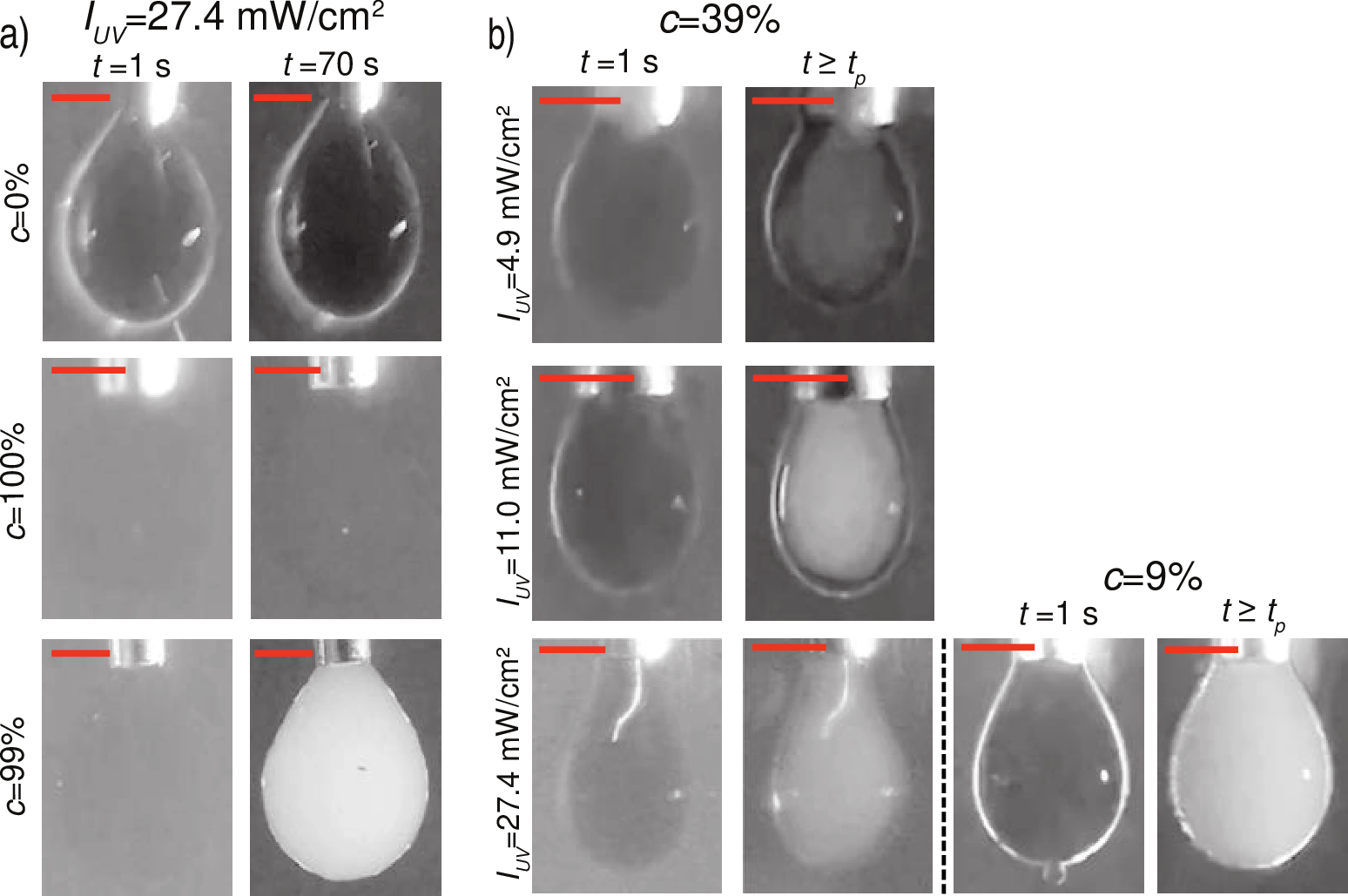}
\caption{Grayscale versions of Fig. \ref{fig: pendant drop images}a in Fig. S5a and Fig. 5c in Fig. S5b. Red bars represent 0.5 mm in length. Fig. S5b Includes additional grayscale pendant drop images for $c=9\%$ exposed to $I_{UV}=27.4$ mW/cm$^2$. }
\label{fig: pendant drop images grayscale}
\end{figure*}

The pendant droplet setup enables the visualization of crosslinking in PEGDA-filled droplets. In the pendant droplet set-up, we record videos of droplets crosslinking over time as they are exposed to different UV intensities. We then convert the videos to grayscale to measure the dynamics of image intensity. Fig. \ref{fig: pendant drop images grayscale} shows grayscale versions of Fig. \ref{fig: pendant drop images}. Additionally, we include a grayscale image of $c=9\%$ exposed to $I_{UV}=27.4$ mW/cm$^2$ at $t=1$ s and $t\ge t_p$.  Droplets with $c=9\%$ exposed to lower UV intensities do not exhibit plateaus in image intensity for the duration of the measurement, which is limited by the capabilities of the UV light source.

\begin{figure*}[h!]
\centering
\includegraphics[width=0.4\columnwidth]{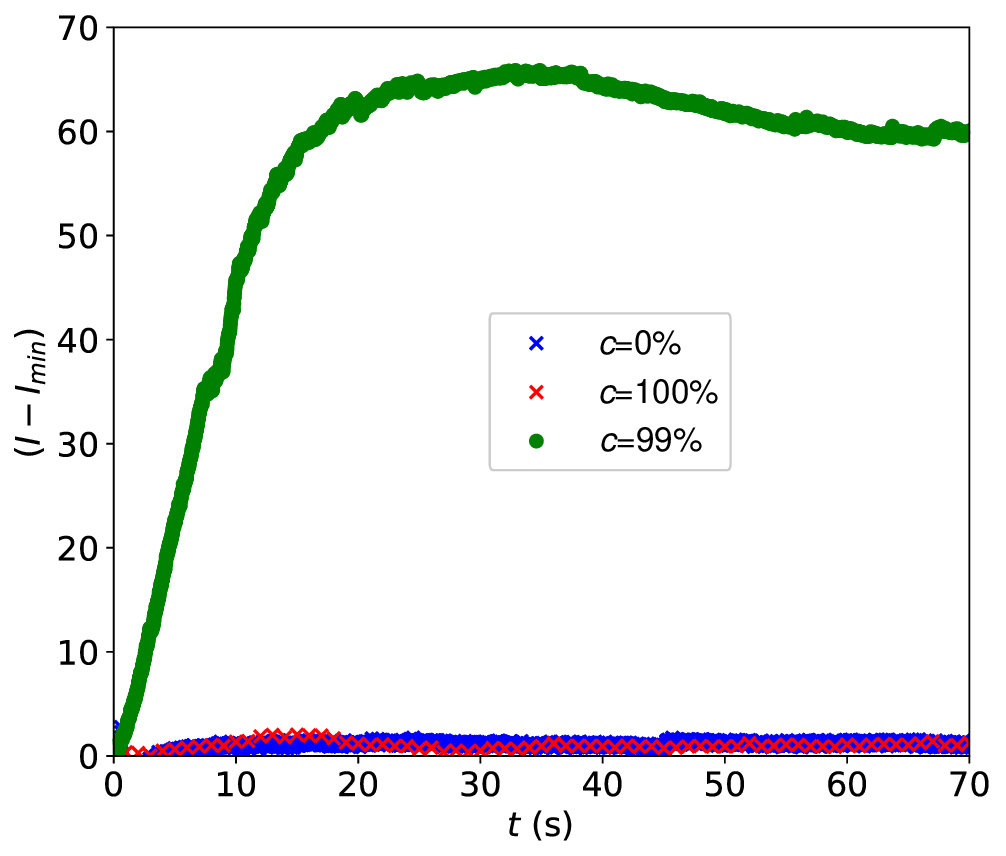}
\caption{Grayscale image intensity ($I$) with the minimum image intensity ($I_{min}$) subtracted, for pendant droplet videos of water (blue), 100\% PEGDA (red), and 99\% PEGDA with 1\% photoinitiator (green). The UV light is turned on at $t=0$s. Image intensity remains relatively constant for $c=0\%$ and $c=100\%$, but increases sharply upon UV illumination and then plateaus for $c=99\%$. } 
\label{fig: control pendant drop}
\end{figure*}

In Fig. \ref{fig: pendant drop images grayscale}a, between $t=1$ s and $t=70$ s, there is a change in droplet image intensity only for $c=99\%$ PEGDA containing 1\% photoinitiator. We confirm the change in image intensity by subtracting the minimum image intensity, $I_{min}$, from the image intensity, $I$. Fig. \ref{fig: control pendant drop} shows $I-I_{min}$ versus time for the conditions in Fig. \ref{fig: pendant drop images grayscale}a. At $c=0\%$ and $c=100\%$ PEGDA, both of which contain no photoinitiator, there is no change in image intensity. For $c=99\%$ PEGDA containing 1\% photoinitiator, the change in image intensity increases and then plateaus, indicating that the change in image intensity is due to PEGDA crosslinking. SI video 2 shows an example of gelation in a droplet filled with $c=39$\% PEGDA exposed to $I_{UV}=11$ mW/cm$^2$.

\par 

\begin{figure*}[h!]
\centering
\includegraphics[width=0.75\columnwidth]{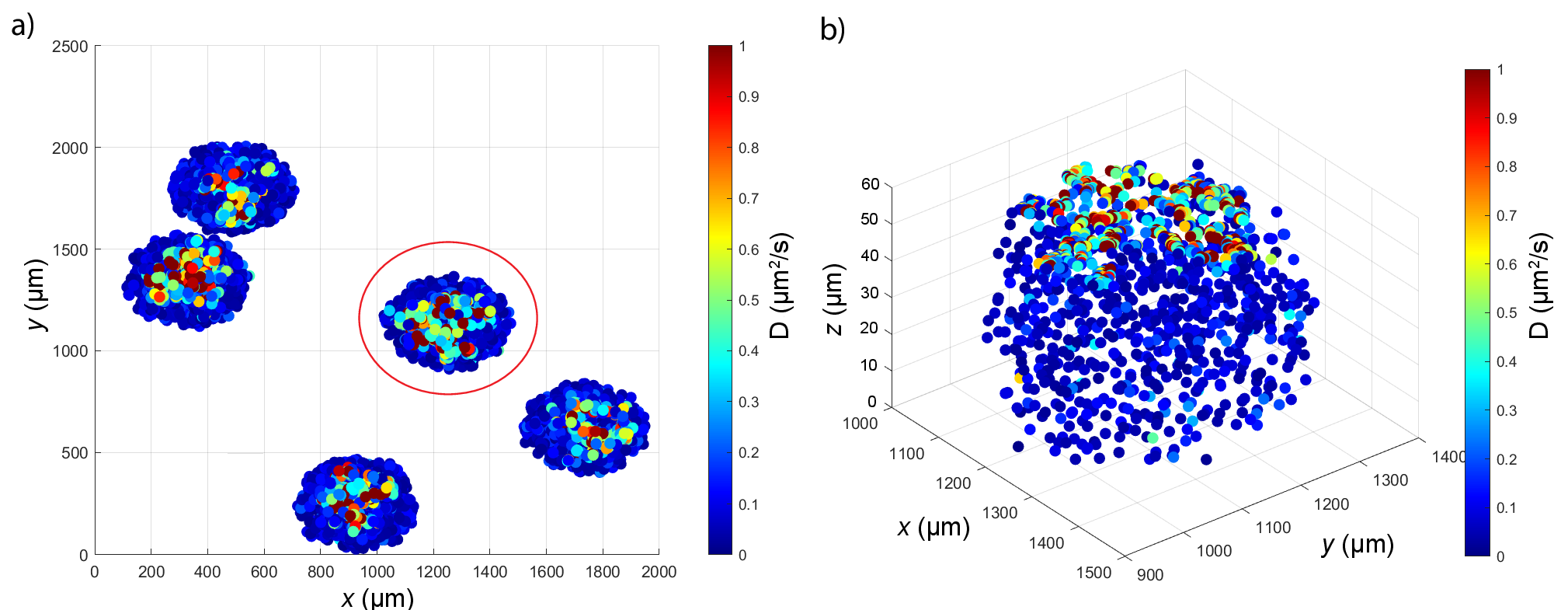}
\caption{Diffusion of 500 nm fluorescent particles in gelled droplets after exposure of $I_{UV}<30$ mW/cm$^2$ for 60 s. Fig. \ref{fig: drop diffusion}a demonstrates that looking at the cured gels from above, we see that the top of the gels have nanoparticles diffusing at various diffusion constants (color bar) from 0 to 1 $\mu m^2/s$. When we focus on just one gel from Fig. \ref{fig: drop diffusion}a (red circle), Fig. \ref{fig: drop diffusion}b, and look at the gelled droplet $z$-stack as well, we see a clearer distinction in nanoparticle diffusion throughout the gel. That is, the top of the gel has a higher diffusion than the rest of the gel.} 
\label{fig: drop diffusion}
\end{figure*}

We corroborate the fluidity of the outer water regime surrounding the inner gel by imaging microparticle laden emulsion drops curing under a microscope.  Fig. \ref{fig: drop diffusion} shows tracer particle diffusivity in large drops that have settled onto a glass slide at the bottom of the water-in-oil emulsion.  The water phase contains fluorescent tracers whose diffusion constants, $\mathcal{D}$, we measure using standard particle tracking methods.  The color bar indicates the diffusion constant in $\mu$m$^2$/s. Fig. \ref{fig: drop diffusion}a shows a top view of five drops in the $x-y$ plane, the plane of the glass slide. The drops each have a diameter $\sim500 \mu$m.  Tracer particles of diameter 500 nm diffusing in water have $\mathcal{D}=1$ $\mu$m$^2$/s. The values of $\mathcal{D}$ span a range of values from 0.1 to 1 $\mu$m$^2$/s.  Fig. \ref{fig: drop diffusion}b shows one of the droplets, indicated in Fig. \ref{fig: drop diffusion}a, using a 3D rendering of the $z$-stack microscopy imaging.  The height of the droplet is only $\sim 60\mu$m, indicating that droplets flatten somewhat due to gravity.  Fig. \ref{fig: drop diffusion}b shows that tracer particles with the largest values of $\mathcal{D}$ sit atop arrested tracer particles.  The arrested particles indicate a gel, which falls to the glass slide because the gel phase is denser than water.  The fluid regime sits on top of the gel, as indicated by greater tracer particle mobility at $z\sim60\mu$m.


\par 

\newpage
\clearpage

\subsection{Oxygen quenching of PEGDA polymerization facilitates the presence of an outer liquid layer}

\begin{figure*}[h!]
\centering
\includegraphics[width=0.5\columnwidth]{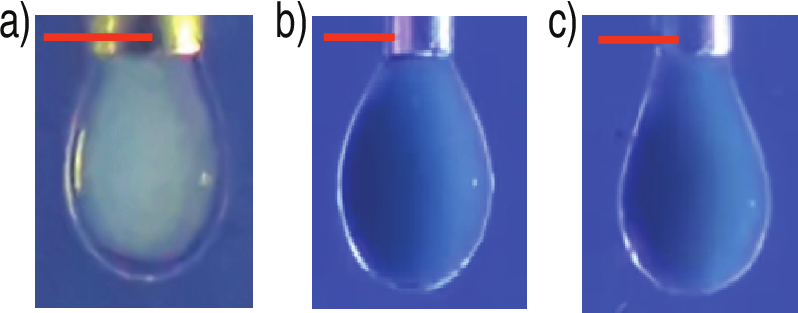}
\caption{\added{Comparisons of pendant droplet images of 39\% PEGDA after exposure of 11 mW/cm$^2$ in different conditions. Fig. \ref{fig: purge}a depicts an image of a fully cured droplet under no nitrogen purge with a liquid halo surrounding the particle. In Fig. \ref{fig: purge}b the mineral oil is purged with nitrogen, while in Fig. \ref{fig: purge}c both the mineral oil and the aqueous PEGDA solution are both purged with nitrogen. The scale bar is 0.5 mm.. }} 
\label{fig: purge}
\end{figure*}

\added{We use the pendant drop setup to test the hypothesis that dissolved oxygen in the mineral oil may diffuse into the drop and quench the free-radical polymerization of PEGDA.  To remove oxygen and create an inert environment, we purge the system with nitrogen before UV exposure.  We image aqueous droplets with $c=39\%$ PEGDA exposed to 11 mW/cm$^2$ after 70s. Fig. \ref{fig: purge} compares images of aqueous droplets in mineral oil with different purging conditions. When the system is not purged with nitrogen, and oxygen remains, a liquid shell surrounds the cured particle, as in \ref{fig: purge}a. When nitrogen is used to purge either the mineral oil solution only or both the mineral oil and aqueous PEGDA solution, the liquid shell disappears. Fig. \ref{fig: purge}b shows a droplet in mineral oil which has been purged with nitrogen.  In Fig. \ref{fig: purge}c both the mineral oil and the aqueous phases have been purged with nitrogen.  The similarity between Fig. \ref{fig: purge}b and c suggests that the oxygen dissolved in mineral oil is sufficient to cause the uncured layer of fluid surrounding the inner gel seen in Fig. \ref{fig: purge}a. In Fig. \ref{fig: purge}b and c, the left side of the drop remains a liquid while the right side of the droplet cures.  This asymmetry is due to the placement of the light guide on the right side of the cuvette that initiates curing from the right side of the droplet.}

\par 

\subsection{Visualizing microscale droplet gelation in flow reveals the presence of uncured water shell and droplet shedding due to constrictions}

\begin{figure*}[h!]
\centering
\includegraphics[width=\columnwidth]{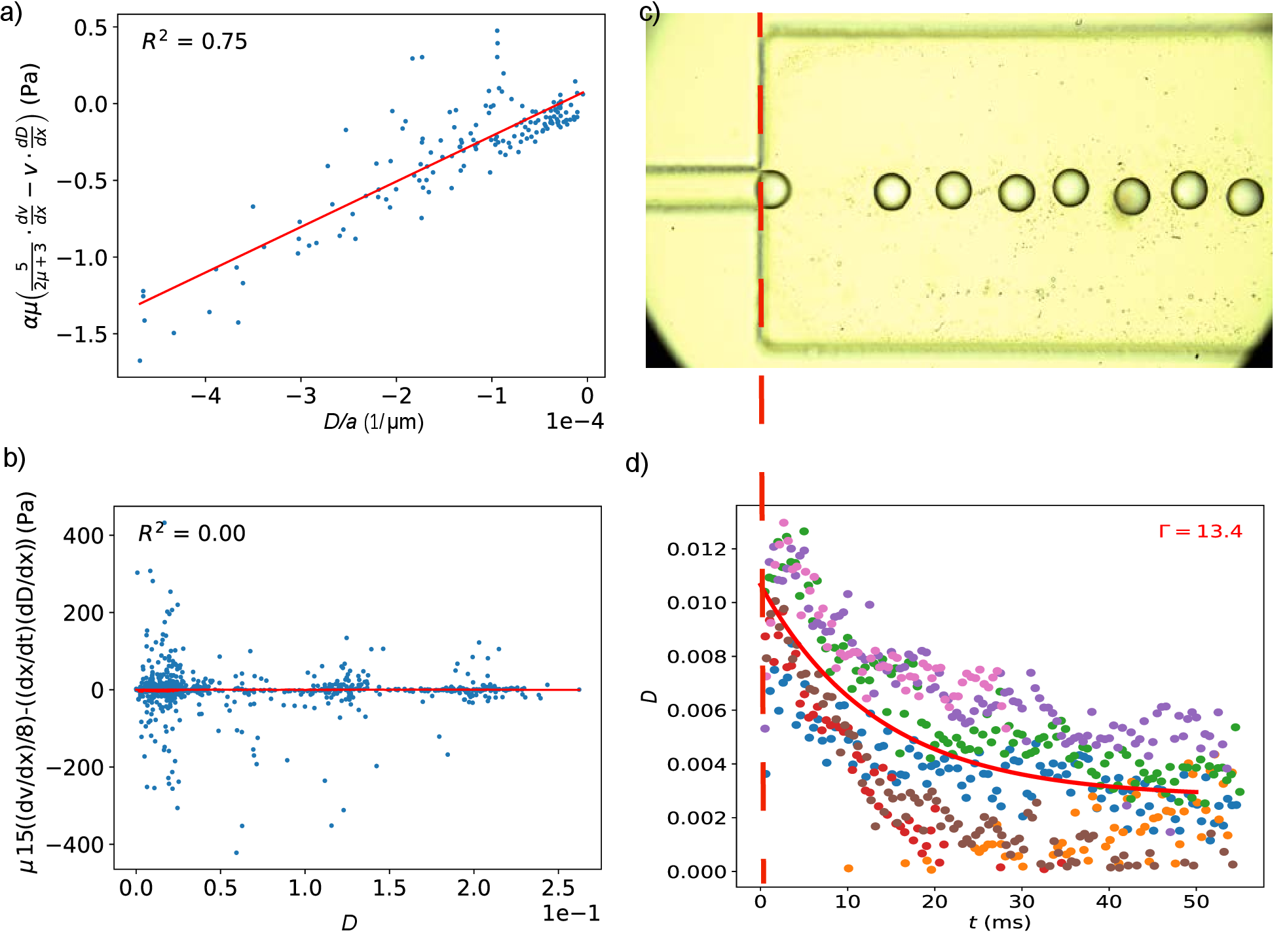}
\caption{Fig. \ref{fig: extension flow SI}a: Taylor fit for the surface tension of droplets, $\alpha \mu (\frac{5}{2\mu+3} \frac{dv}{dx} - v\frac{dD}{dx})$ vs $D/a$. Slope yields $\sigma$. Fig. \ref{fig: extension flow SI}b: Fit for the elastic modulus of particles, $\frac{15}{8}\mu(\frac{dv}{dx}-\frac{dx}{dt}\frac{dD}{dx})$ vs $D$. Slope yields $E$. Fig. \ref{fig: extension flow SI}c: Example of $c=19\%$ particles moving through a constriction. The red vertical line indicates $x=0$. Fig. \ref{fig: extension flow SI}d: $D$ vs $t$ of the example presented in Fig. \ref{fig: extension flow SI}c. The red fit represents fit to the equation: $D=Ae^{-\frac{\Gamma}{t}}+C$ where $A$ and $C$ are fitting constants. $\Gamma$ is a relaxation time.  } 
\label{fig: extension flow SI}
\end{figure*}

In a device such as the one shown in Fig. \ref{UVschematic3}b, we measure relaxation dynamics of droplets and particles back to their original shape as they leaves a constriction. We extract surface tension by taking the slope of $\alpha \mu (\frac{5}{2\mu+3} \frac{dv}{dx} - v\frac{dD}{dx})$ versus $D/a$, following established methods. \cite{Hudson2005, chen.dutcher.2020, Shaulsky2025} An example of this analysis is shown in Fig. \ref{fig: extension flow SI}a. The example of $c=9\%$ PEGDA droplets with no UV exposure yields a surface tension of $3.0$ mN/m and an $R^2$ fit of 0.75. For elastic modulus of particles, we measure the slope of  $\frac{15}{8}\mu(\frac{dv}{dx}-\frac{dx}{dt}\frac{dD}{dx})$ vs $D$, following established methods \cite{Villone2019}. The example shown in Fig. \ref{fig: extension flow SI}a reveals that the behavior of particles exiting the constriction does not follow the dynamics expected for uniformly elastic particles ($R^2=0$), which may indicate they are too stiff to exhibit significant relaxation. Fig. \ref{fig: extension flow SI}c and d provide an alternate method for characterizing particles using an example of $c=19\%$ PEGDA exposed to 4600 mJ/cm$^2$. The vertical red line in Fig.\ref{fig: extension flow SI}c and d indicates $t=0$s as particles exit the constriction. $D$ is measured as a function of $t$ and then fit to $D=Ae^{-\Gamma/t}+C$ where $A$ and $C$ are constants and $\Gamma$ is a relaxation time. SI video 3 shows a compilation of 6 different experimental conditions explored in tight constriction flow, as indicated in table \ref{tab: transient table}, ranging from uncured to fully cured. At time points 0:05 and 0:12 in the video, the separation of uncured water from the cured core of the particle is visible. Later in the video, even with more fully cured particles, evidence of an external water film can be seen immediately after the particles exit the constriction, for instance at time points 0:26 and 0:30, as some water adheres to the wall of the constriction near the exit.

\begin{figure*}[h!]
\centering
\includegraphics[width=0.5\columnwidth]{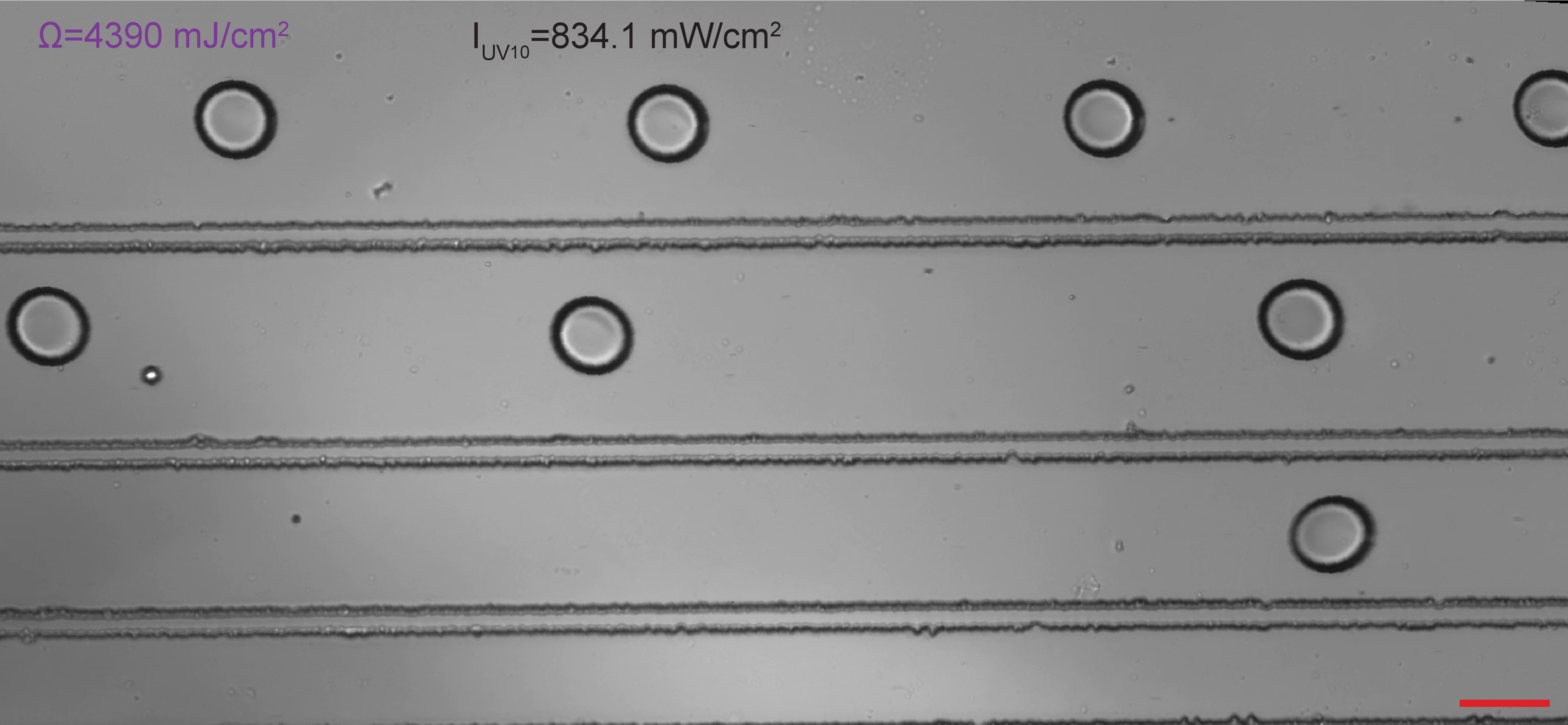}
\caption{$c=19\%$ PEGDA particles in the steady device exposed to 4390 mJ/cm$^2$ and 834.1 mW/cm$^2$. The red scale bar represents 80 $\mu$m.} 
\label{fig: 19 steady}
\end{figure*}

Given the presence of uncured water around an inner gel observed in capillary micromechanics, pendant drop crosslinking, and flow through tight fluidic constrictions, Fig. \ref{fig: 19 steady} shows an image of a sample with $c=19\%$ exposed to $\Omega=4390$ mJ/cm$^2$ and $I_{UV_{10}}=834.1$ mW/cm$^2$.  Within the dark outer contour of each droplet or particle, we see a lighter outer shell surrounding a darker inner shadow.  This image corresponds to a measurement of $\tau$ in Fig. \ref{fig: restoring stress} that is suggestive of the surface tension of surfactant-covered PEGDA-filled water drops in oil.



